\documentclass[aps,prd,onecolumn,superscriptaddress,notitlepage]{revtex4-1}

\usepackage{aas_macros}
\usepackage[utf8]{inputenc}
\usepackage[T1]{fontenc}
\usepackage[english,francais]{babel}
\usepackage{mathrsfs}
\usepackage{amsmath}
\usepackage{amssymb}
\usepackage{tipa}
\usepackage{ntheorem}
\usepackage{braket}
\usepackage{graphicx}
\usepackage{bm}
\usepackage{wasysym}
\usepackage{enumitem}
\usepackage{stmaryrd}
\usepackage[breaklinks=true,colorlinks,citecolor=blue,linkcolor=blue,urlcolor=blue]{hyperref}
\usepackage{tikz}
\usetikzlibrary[patterns]
\usetikzlibrary{shapes}

\theoremseparator{.}

\makeatletter
\newskip\@bigflushglue \@bigflushglue = -100pt plus 1fil

\def\bigcentering{\let\\\@centercr\rightskip\@bigflushglue
\leftskip\@bigflushglue
\parindent\z@\parfillskip\z@skip}

\makeatother

\allowdisplaybreaks[1]

\newcommand{\dd}{\mathrm{d}}

\begin{document}

\title{Relativistic modeling of atmospheric occultations with time transfer functions}
\author{A. Bourgoin}
\email{adrien.bourgoin@unibo.it}
\affiliation{Dipartimento di Ingegneria Industriale, Alma Mater Studiorum -- Università di Bologna, Via Fontanelle 40, 47121 Forlì, Italy}
\affiliation{Interdepartmental Center for Industrial Research in Aerospace (CIRI AERO), Alma Mater Studiorum -- Università di Bologna, Via B. Carnaccini 12, 47121 Forlì, Italy}
\author{M. Zannoni}
\affiliation{Dipartimento di Ingegneria Industriale, Alma Mater Studiorum -- Università di Bologna, Via Fontanelle 40, 47121 Forlì, Italy}
\affiliation{Interdepartmental Center for Industrial Research in Aerospace (CIRI AERO), Alma Mater Studiorum -- Università di Bologna, Via B. Carnaccini 12, 47121 Forlì, Italy}
\email{adrien.bourgoin@unibo.it}
\author{L. Gomez Casajus}
\affiliation{Dipartimento di Ingegneria Industriale, Alma Mater Studiorum -- Università di Bologna, Via Fontanelle 40, 47121 Forlì, Italy}
\affiliation{Interdepartmental Center for Industrial Research in Aerospace (CIRI AERO), Alma Mater Studiorum -- Università di Bologna, Via B. Carnaccini 12, 47121 Forlì, Italy}
\email{adrien.bourgoin@unibo.it}
\author{P. Tortora}
\affiliation{Dipartimento di Ingegneria Industriale, Alma Mater Studiorum -- Università di Bologna, Via Fontanelle 40, 47121 Forlì, Italy}
\affiliation{Interdepartmental Center for Industrial Research in Aerospace (CIRI AERO), Alma Mater Studiorum -- Università di Bologna, Via B. Carnaccini 12, 47121 Forlì, Italy}
\email{adrien.bourgoin@unibo.it}
\author{P. Teyssandier}
\affiliation{SYRTE, Observatoire de Paris, Universit\'e PSL, CNRS, Sorbonne Universit\'e, LNE, 61 avenue de l'Observatoire, 75014 Paris, France}


\begin{abstract}
  \emph{Context. }Occultation experiments represent unique opportunities for probing remotely physical properties of atmospheres. The data processing requires one to properly account for refractivity while modeling the time/frequency transfers of an electromagnetic signal. On theoretical grounds, little work have been done concerning the elaboration of a covariant approach for modeling occultation data.

  \emph{Aims. }We present an original method allowing one to derive up to the appropriate order fully analytical expressions for the covariant description of time/frequency transfers during an atmospheric occultation experiment.

  \emph{Methods. }We make use of two independent powerful relativistic theoretical tools, namely the optical spacetime metric, and the time transfer functions formalism. The first one allows us to consider refractivity as spacetime curvature while the second one is used to determine the time/frequency transfers occurring in a curved spacetime.

  \emph{Results. }We provide the integral form of the time transfer function up to any post-Minkowskian order. We specify the discussion to a stationary optical spacetime describing an occultation by a steady rotating and spherically symmetric atmosphere. Explicit analytical expressions for the time/frequency transfers are provided at the first post-Minkowskian order and their accuracy is assessed by comparing them to results of a numerical integration of the equations for optical rays.

  \emph{Conclusions. }The method accurately describes vertical temperature gradients and properly accounts for light-dragging effect due to the motion of the optical medium. It can be pushed further in order to derive the explicit form of the time transfer function at higher order and beyond the spherical symmetry assumption.
\end{abstract}

\maketitle

\section{Introduction}

Theoretical problems dealing with time and frequency transfers require one to know the function relating the (coordinate) time transfer to the coordinate time at reception and to the spatial coordinates of the reception and emission point-events. Such a function is called a reception time transfer function. Obviously, an emission time transfer function can be introduced too. The formalism which aims at determining the time transfer functions is called the time transfer functions formalism. It was first introduced by \citet{2002PhRvD..66b4045L} relying on the theory of the world function developed by \citet{SyngeBookGR}. Later on a general post-Minkowskian expansion of the world and the time transfer functions was proposed by \citet{2004CQGra..21.4463L} and then the method was refined by \citet{2008CQGra..25n5020T} thanks to a simplified iterative procedure. The time transfer functions formalism presents the great advantage to spare the trouble of integrating the geodesic equation which usually leads to heavy calculations especially beyond the post-Minkowskian regime \citep{1983PhRvD..28.3007R,1987KFNT....3....8B}. Until recently, the time transfer functions formalism was systematically applied to the physical spacetime considering only gravitational effects on a ray of light propagating in a vacuum. However, \citet{PhysRevD.101.064035} showed that theoretical problems dealing with optical rays propagating into flowing dielectrics medium could also be solved making use of the time transfer functions formalism by means of a powerful theoretical tool known as the optical metric or the Gordon's metric of spacetime.

When a ray of light is propagating into an optical medium its trajectory does not follow a null geodesic path of the physical spacetime since light and matter interacts. Conventionally, the real trajectory is thus determined by solving Maxwell's equation within the framework of geometrical optics. However, an other interesting possibility initially proposed by \citet{doi101002andp19233772202} is to introduce an artificial optical spacetime which implicitly accounts for the interaction between light and matter such that optical rays follow null geodesics of that new optical spacetime. In other words, refractivity is treated as spacetime curvature in the optical spacetime. Therefore, the time and frequency transfers can still be determined with the time transfer functions formalism even when considering a ray of light crossing through an optical medium.

Occultation experiments are an example of observing technique requiring a careful treatment of refractivity while modeling the time/frequency transfers. The method consists in measuring remotely the physical properties of a planetary atmosphere while the source of an electromagnetic signal is being occulted by the atmosphere. When the source is a radio signal emitted by a spacecraft's antenna the experiment is called an atmospheric radio occultation \citep{1965Sci...149.1243K,1965JGR....70.3217F,1968P&SS...16.1035F,1985AJ.....90.1136L,1987JGR....9214987L,1992AJ....103..967L,2012Icar..221.1020S,2015RaSc...50..712S} whereas it is called an atmospheric stellar occultation when the source is made of visible or near-infrared light emitted by a distant star \citep{1994A&A...288..985R,2006JGRE..11111S91S}. In practice, two methods are usually employed for processing occultations data, namely the Abel inversion for spherical symmetry \citep{1968JGR....73.1819P,1999AnGeo..17..122S} and the numerical ray-tracing for any generic cases \citep{2015RaSc...50..712S}. The former is an exact expression providing the index of refraction profile directly from the bending angle which is itself retrieved from the frequency transfer. The latter consists in a numerical integration of the equations for optical rays across a layered atmosphere. The refractivity in each layer and the initial pointing direction are iteratively determined such that the computed frequency coincides with the observed frequency. While the numerical ray-tracing method is the most general one, it does not provide a comprehensive description of the light path and requires a high computational time. On the other hand, if the Abel inversion method does not require to numerically solve for the equations for optical rays, it can only be applied to spherically symmetric atmospheres and cannot account for the atmospheric time delay which eventually affects the determination of the emitter's position (i.e. the spacecraft in a one-way downlink configuration during a radio occultation event). In addition, such as the numerical ray-tracing, the Abel inversion does not provide a comprehensive description of the light path. An analytical expression describing the time/frequency transfers for radio occultation experiments would allow to overcome these issues.

With this goal in mind, \citet{2019A&A...624A..41B} proposed a new approach exploiting similarities between equations of geometrical optics and equations of celestial mechanics. It consists in expressing the equations of geometrical optics into a set of first order perturbation equations (similar to Gauss equations of celestial mechanics) which are better suited for finding analytical solutions beyond the spherical symmetry assumption. However, even if the perturbation equations can be solved more easily than the equations of geometrical optics it is still a challenging task to solve for the second order expressions or to incorporate the light-dragging effect caused by the motion of the optical medium.

In this paper, we make use of the time transfer functions formalism considering an optical spacetime in order to model accurately the time/frequency transfers for occultation experiments involving a flowing spherically symmetric atmosphere. The determination of the time transfer functions allows to account for the atmospheric time delay. In addition, the formalism being fully covariant, it naturally accounts for the light-dragging effect due to the motion of the optical medium and constitutes in that sense a significant improvement with respect to the perturbation equations approach. 

The paper is organized as follows. Sec. \ref{sec:not} lists the notations and assumptions we make throughout the paper. Sec.~\ref{sec:relopt} recalls some basics about relativistic geometrical optics and allows us to define the frequency, the index of refraction, the refractivity, and the optical metric. The equations for optical rays propagating in an isotropic dispersive medium are derived in the same section. The reader who is already familiar with relativistic geometrical optics can skip this section. Sec.~\ref{sec:TF} recalls basics about the time transfer functions formalism and introduces the refractive delay function and the post-Minkowskian parameter $N_0$. The integral form of the delay function is given at any post-Minkowskian order. Sec.~\ref{sec:app} is an application to occultation experiments involving steady rotating and spherically symmetric atmospheres. The refractivity profile is built in Sec.~\ref{sec:mathmod} considering an exponential pressure profile and a polynomial temperature profile of arbitrary degree. The refractive delay function is finally solved at first post-Minkowskian order in the limit where the angular velocity of the optical medium is small with respect to the speed of light in a vacuum. The expressions for the time/frequency transfers are given explicitly. Sec. \ref{sec:NumRT} assesses the accuracy of the first order solutions for the time/frequency transfers by comparing them to results of a numerical integration of the equations for optical rays propagating into a nondispersive isotropic medium. Finally, we give our conclusions in Sec.~\ref{sec:ccl}. Sec. \ref{sec:Abel} is a discussion about the Abel transform method for retrieving the refractivity from the frequency transfer while considering the light-dragging effect.

\section{General assumptions and notations}
\label{sec:not}

The influence of gravity on the propagation of light is regarded as negligible, so the physical metric $g$ of spacetime is assumed to be a Minkowski metric. Greek indices run from 0 to 3, Latin indices run from 1 to 3. We systematically make use of an orthonormal Cartesian coordinate system $(x^\mu)=(x^0,x^i)$, so the components of the physical metric may be written as
\begin{equation}
  g_{\mu\nu}=\eta_{\mu\nu}\text{,}
  \label{eq:flat}
\end{equation}
where
\begin{equation}
  \eta_{\mu\nu}=\text{diag}(+1,-1,-1,-1)\text{.}
\end{equation}

We put $x^0=ct$, with $c$ being the speed of light in a vacuum and $t$ a time coordinate, and we denote by $\bm x$ the triple of spatial coordinates $(x^1,x^2,x^3)$. More generally, we use the notation $\bm a=(a^i)=(a^1,a^2,a^3)$ for a triple constituted by the spatial components of a 4-vector, and $\underbar{$\bm b$}=(b_i)=(b_1,b_2,b_3)$ for a triple built with the spatial components of a covariant 4-vector.

Given the triples $\bm a$, $\bm b$, and $\underbar{$\bm c$}$, the usual Euclidean scalar product $\bm a\cdot\bm b$ is denoted by $a^ib^i=\delta_{ik}a^ib^k$, where $\delta_{ik}$ is the Kronecker delta. Similarly, $\bm a\cdot\underbar{$\bm c$}$ denotes the quantity $a^ic_i=\delta_{ik}a^ic_k$. In each case, Einstein's summation convention on repeated indices is used. Furthermore, $\Vert\bm a\Vert$ denotes the Euclidean norm of $\bm{a}$: $\Vert\bm{a}\Vert=\sqrt{\bm a\cdot\bm a}$. Similarly, $\Vert\underbar{$\bm c$}\Vert$ denotes the quantity $\Vert\underbar{$\bm c$}\Vert=\sqrt{\underbar{$\bm c$}\cdot\underbar{$\bm c$}}$.

For the sake of legibility, we employ $(f)_{x}$ or $[f]_{x}$ instead of $f(x)$ whenever necessary. When a quantity $f(x)$ is to be evaluated at two point-events $x_A$ and $x_B$, we employ $(f)_{A/B}$ to denote $f(x_A)$ and $f(x_B)$, respectively. The partial differentiation of $f$ with respect to $x^\mu$ is denoted by $\partial_\mu f$ or by $f_{,\mu}$.

\section{Relativistic geometrical optics}
\label{sec:relopt}

This paper is devoted to the propagation of light rays through a linear, isotropic, and nondispersive medium filling a spatially bounded region $\mathcal D$ of spacetime. The regions of spacetime outside $\mathcal D$ are supposed to be empty of any matter.

The electromagnetic properties of the medium are characterized by two scalar functions, the permittivity $\epsilon(x)$ and the permeability $\mu(x)$. The index of refraction of the medium is the scalar function defined by the well-known relationship
\begin{equation}
  n(x)=c\sqrt{\epsilon(x)\mu(x)}\text{.}
  \label{eq:defniso}
\end{equation}
Moreover, it is assumed that the medium is made of a fluid schematized by a flow of particles which are not colliding. The unit 4-velocity vector of a particle of the fluid at a point-event $x$ of its world-line is denoted by $w^\mu(x)$. Outside $\mathcal D$, the permittivity and the permeability reduce to their vacuum values $\epsilon(x)=\epsilon_0$ and $\mu(x)=\mu_0$, respectively. Since $c=(\epsilon_0\mu_0)^{-1/2}$, the index of refraction then reduces to $n(x)=1$. The expression for the refractivity is obtained by subtracting its vacuum value to the index of refraction, namely
\begin{equation}
  N(x)=n(x)-1\text{.}
  \label{eq:defN}
\end{equation}

In the context of the geometrical optics approximation, the light rays propagating through our medium are the bicharacteristic curves of the so-called eikonal equation, which reads as
\begin{equation}
  \bar g^{\mu\nu}\partial_\mu\mathscr S\partial_\nu\mathscr S=0\text{,}
  \label{eq:eik}
\end{equation}
where $\mathscr S(x)$ is the eikonal function and $\bar g^{\mu\nu}$ is the contravariant tensor defined by
\begin{equation}
  \bar g^{\mu\nu}=g^{\mu\nu}+\kappa^{\mu\nu}\text{,} \qquad \kappa^{\mu\nu}= (n^2-1)w^\mu w^\nu\text{.}
  \label{eq:Gorcon}
\end{equation}
Let us denote by $\bar g_{\mu\nu}$ the quantities such that
\begin{equation}
  \bar g_{\mu\alpha}\bar g^{\alpha\nu}=\delta_\mu^{\ \nu}\text{.}
  \label{eq:inv}
\end{equation}
An elementary calculation leads to
\begin{equation}
  \bar g_{\mu\nu}=g_{\mu\nu}+\gamma_{\mu\nu}\text{,} \qquad \gamma_{\mu\nu}=-\left(1-\frac{1}{n^2}\right)w_\mu w_\nu\text{.}\label{eq:Gorcov}
\end{equation}

The quantities $\bar g_{\mu\nu}$ can be regarded as the components of a Lorentzian metric $\bar g$ defined on the region $\mathcal D$. This new metric is called the optical metric associated with the refracting medium, a terminology justified by the following considerations, which have been previously derived by \citet{doi101002andp19233772202,1957ArRMA...1...54Q}, and \citet{1967ZNatA..22.1328E}. Let us define the 4-wave covector field $k_{\mu}$ as
\begin{equation}
  k_{\mu}(x)=\partial_\mu\mathscr{S}(x)\text{.}
  \label{eq:kcovdef}
\end{equation}
A contravariant vector field $k^\mu$ can be associated with this covector by putting
\begin{equation}
  k^\mu=g^{\mu\nu}k_\nu\text{.}
  \label{eq:kcondef}
\end{equation}
The light rays $x^\mu=x^\mu(\zeta)$ associated with a solution $\mathscr S$ of the eikonal equation \eqref{eq:eik} are the integral curves of the contravariant vector field $\bar k^\mu$, i.e. are solutions of the differential system \citep{perlick2000ray}
\begin{equation}
  \frac{\dd x^\mu}{\dd\zeta}=\bar g^{\mu\nu}\partial_\nu\mathscr{S}\big(x(\zeta)\big)\text{.}
  \label{eq:tg}
\end{equation}

A classical calculation shows that the solutions of \eqref{eq:tg} are null geodesics of the optical metric $\bar g$ \citep{SyngeBookGR,perlick2000ray}, $\zeta$ being an affine parameter. The null character of the rays is directly inferred from \eqref{eq:inv}, \eqref{eq:tg}, and \eqref{eq:eik}. We have indeed:
\begin{equation}
  \bar g_{\mu\nu}\frac{\dd x^\mu}{\dd\zeta}\frac{\dd x^\nu}{\dd\zeta}=\bar g^{\alpha\beta}\partial_\alpha\mathscr S\partial_\beta\mathscr S=0\text{.}
\end{equation}

The eikonal equation \eqref{eq:eik} is the Jacobi equation associated with the Hamiltonian
\begin{equation}
  H(x^\mu,k_\nu)=\frac{1}{2} \bar g^{\alpha\beta}(x)k_\alpha k_\beta\text{,}
  \label{eq:Hambg}
\end{equation}
where $k_\nu$ must be regarded as conjugate canonical variables of $x^\mu$. As a consequence, the light rays in the region $\mathcal D$ can be considered as solutions of the set of canonical equations \citep{1957ArRMA...1...54Q}
\begin{subequations}\label{eq:RTdefgen}
\begin{align}
  \frac{\dd x^\mu}{\dd \zeta}&=\frac{\partial H}{\partial k_\mu}=k^\mu+(n^2-1)(w^\nu k_\nu)w^\mu\text{,}\label{eq:RTdefgen1}\\
  \frac{\dd k_\mu}{\dd \zeta}&=-\frac{\partial H}{\partial x^\mu}=-nn_{,\mu}(w^\nu k_\nu)^2-(n^2-1)(w^\nu k_\nu)w^\alpha_{\ ,\mu}k_\alpha\text{,}
\end{align}
\end{subequations}

It must be noted that the eikonal function is constant along a light ray. Indeed, it follows from \eqref{eq:tg} that
\begin{equation}
  \frac{\dd\mathscr S}{\dd\zeta}=\frac{\dd x^\mu}{\dd \zeta}\partial_\mu\mathscr S=\bar g^{\mu\nu}\partial_\mu\mathscr S\partial_\nu\mathscr S=0\text{.}
  \label{eq:Scte}
\end{equation}
This property is at the core of the procedure developed in the next section.

For numerical integration, it is relevant to separate space and time components in \eqref{eq:RTdefgen}. Let $l_i$ be the components defined by
\begin{equation}
  l_i=\frac{k_i}{k_0}\text{,}
  \label{eq:defl}
\end{equation}
and let $\ell$ be a new parametrization of the curves for optical rays \citep{2015RaSc...50..712S} such that 
\begin{equation}
  \dd\ell=nk_0\dd\zeta\text{.}
\end{equation}
After inserting these new quantities into the set of canonical equations \eqref{eq:RTdefgen}, we find for the time components:
\begin{subequations}\label{eq:RTdef}
\begin{align}
  \frac{\dd x^0}{\dd\ell}&=\frac{1}{n}\left[1+(n^2-1)\left(w^0+w^kl_k\right)w^0\right]\text{,}\\
  \frac{\dd\ln\Vert k_0\Vert}{\dd\ell}&=-n_{,0}\left(w^0+w^kl_k\right)^2-\frac{(n^2-1)}{n}\left(w^0+w^kl_k\right)\left(w^0_{\ ,0}+w^k_{\ ,0}l_k\right)\text{,}\label{eq:RTdefk0}
\end{align}
and, for the space components:
\begin{align}
  \frac{\dd x^i}{\dd\ell}&=-\frac{1}{n}\left[l_i-(n^2-1)\left(w^0+w^kl_k\right)w^i\right]\text{,}\\
  \frac{\dd l_i}{\dd\ell}&=-n_{,i}\left(w^0+w^kl_k\right)^2-\frac{(n^2-1)}{n}\left(w^0+w^kl_k\right)\left(w^0_{\ ,i}+w^k_{\ ,i}l_k\right)-\frac{\dd\ln\Vert k_0\Vert}{\dd\ell}l_i\text{.}
\end{align}
\end{subequations}
Equations \eqref{eq:RTdef} may be particularly interesting in the case where the index of refraction $n$ and the unit 4-velocity $w^\mu$ do not depend on time, and we use them in our attempt of numerical integration (see Sec. \ref{sec:NumRT}). It is worthy of note that in this case the component $k_0$ is constant along each light ray.

\section{Refractive delay function}
\label{sec:TF}

In this section, we introduce the formalism of time transfer functions that we apply to optical spacetime in order to describe refraction due to a linear, isotropic, and nondispersive medium.

\subsection{Time transfer functions formalism}
\label{sec:TTF}

\begin{figure}
  \begin{center}
    \includegraphics[scale=0.18]{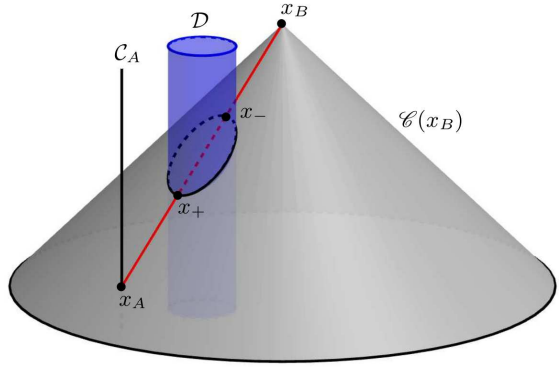}
  \end{center}
  \caption{Illustration of an occultation event in spacetime. The past light cone $\mathscr{C}(x_B)$ of the reception point-event $x_B$ intersects $\mathcal C_A$ the world-line of the emitter at the point-event $x_A$. The zeroth-order null light path (red line) joining the emission point-event $x_A$ to the reception point-event $x_B$ lies on the surface of $\mathscr{C}(x_B)$. The domain $\mathcal D$ represents the limit of the refractive region while $x_-$ and $x_+$ are the intersection point-events between $\mathcal D$ and the zeroth-order null light path. The path of integration (dashed red line) is limited to the portion of the zeroth-order null light path crossing through $\mathcal D$.}
  \label{fig:lightcone}
\end{figure}

Let us consider a light ray $\Gamma_{AB}$ starting from an emission point-event $(x_A^0,\bm x_A)$ and arriving at a reception point-event $(x_B^0,\bm x_B)$. We suppose that a part of $\Gamma_{AB}$ travels through the domain $\mathcal D$, while the other part travels through a vacuum, that is a medium such that $n=1$ (see Fig.~\ref{fig:lightcone}). According to \eqref{eq:Scte}, the phase is a first integral along $\Gamma_{AB}$ so we have a relation as follows
\begin{equation}
  \mathscr S\big(x_B^0,\bm x_B\big)-\mathscr S\big(x_A^0,\bm x_A\big)=0\text{.}
  \label{eq:SASB}
\end{equation}
Equation \eqref{eq:SASB} shows that $x_A^0$ is an implicit function of $\bm x_A$, $x_B^0$, and $\bm x_B$. Hence, it is appropriate to introduce $\mathcal{T}$ the reception time transfer function associated with $\Gamma_{AB}$ as
\begin{equation}
  x_B^0-x_A^0=c\mathcal{T}\big(\bm x_A,x_B^0,\bm x_B\big)\text{.}
  \label{eq:TTFdef}
\end{equation}
Obviously, an emission time transfer function can be introduced too. Hereafter, we merely consider the case at reception which is better suited for a downlink one-way transfer during an occultation event with a radio signal that is recorded when received.

A relevant theorem about the components of the 4-wave covector can be directly inferred from \eqref{eq:SASB} and \eqref{eq:TTFdef} (see \citet{2004CQGra..21.4463L}). Indeed after writing \eqref{eq:TTFdef} in the equivalent form: $x_A^0=x_B^0-c\mathcal{T}(\bm x_A,x_B^0,\bm x_B)$, and then inserting this relation into \eqref{eq:SASB}, we get an expression as follows
\begin{equation}
  \mathscr S\big(x_B^0,\bm x_B\big)-\mathscr S\big(x_B^0-c\mathcal{T}\big(\bm x_A,x_B^0,\bm x_B\big),\bm x_A\big)=0\text{.}
  \label{eq:SASBR}
\end{equation}
Equation \eqref{eq:SASBR} is in fact an identity. So straightforward differentiations of this last relation with respect to $\bm x_A$, $x_B^0$, and $\bm x_B$ lead to the following set of equations:
\begin{subequations}\label{eq:partSASB}
\begin{align}
  &\frac{\partial\mathscr S}{\partial x_A^i}-\frac{\partial\mathscr S}{\partial x_A^0}\frac{\partial\mathcal R}{\partial x_A^i}=0\text{,}\\
  &\frac{\partial\mathscr S}{\partial x_B^0}-\frac{\partial\mathscr S}{\partial x_A^0}\left(1-\frac{\partial\mathcal R}{\partial x_B^0}\right)=0\text{,}\\
  &\frac{\partial\mathscr S}{\partial x_B^i}+\frac{\partial\mathscr S}{\partial x_A^0}\frac{\partial\mathcal R}{\partial x_B^i}=0\text{.}
\end{align}
\end{subequations}
where we introduced $\mathcal R$ a reception range transfer function as
\begin{equation}
  \mathcal{R}(\bm x_A,x_B)=c\mathcal{T}\big(\bm x_A,x_B^0,\bm x_B\big)\text{.}
  \label{eq:RTFrdef}
\end{equation}
Using Eqs. \eqref{eq:kcovdef} and \eqref{eq:defl}, it is easily seen that Eqs. \eqref{eq:partSASB} imply the following relationships:
\begin{subequations}\label{eq:kTFdef}
\begin{align}
  (l_i)_A&=\frac{\partial\mathcal{R}}{\partial x_A^i}\text{,}\label{eq:kTFA}\\
  (l_i)_B&=-\frac{\partial\mathcal{R}}{\partial x_B^i}\left(1-\frac{\partial\mathcal{R}}{\partial x_B^0}\right)^{-1}\text{,}\label{eq:kTFB}
\end{align}
and
\begin{equation}
  \frac{(k_0)_B}{(k_0)_A}=1-\frac{\partial\mathcal{R}}{\partial x_B^0}\text{.}
  \label{eq:kTF0}
\end{equation}
\end{subequations}
Unsurprisingly, these relations are similar to Eqs. (40)-(42) of \citet{2004CQGra..21.4463L} established for optical rays in a vacuum. We actually see that they are still valid for optical rays propagating into a linear, isotropic, and nondispersive medium within the framework of geometrical optics.

The main interest of \eqref{eq:kTFdef} lies in the fact that they enable us to calculate the Doppler effect between an emitter and a receiver when the explicit expression of the function $\mathcal R$ is known. Indeed, it is well known \citep{SyngeBookGR,2001A&A...370..320B} that the Doppler frequency shift measured between an emitter and a receiver can be expressed as
\begin{equation}
  \frac{\nu_B}{\nu_A}=\frac{(u^{\mu}k_\mu)_B}{(u^{\mu}k_\mu)_A}=\frac{(u^0k_0)_B}{(u^0k_0)_A}\frac{(1+\beta^il_i)_B}{(1+\beta^il_i)_A}\text{,}
  \label{eq:dop}
\end{equation}\\
where $(u^\mu)_{A/B}$ are the emitter/receiver's unit 4-velocity vectors defined by
\begin{equation}
  (u^\mu)_{A/B}=\left(\frac{\dd x^\mu}{\dd s}\right)_{A/B}\text{,}
  \label{eq:4veldef}
\end{equation}
with
\begin{equation}
  \dd s^2=g_{\mu\nu}\dd x^\mu\dd x^\nu\text{.}
  \label{eq:ds}
\end{equation}
The unit 4-velocity is by definition a unit vector for the physical metric of spacetime \eqref{eq:flat}, hence
\begin{equation}
  (u^0)_{A/B}=\left(\frac{1}{\sqrt{1-\Vert\bm\beta\Vert^2}}\right)_{A/B}\text{,}
  \label{eq:4vel}
\end{equation}
where $(\beta^i)_{A/B}$ denote the coordinate 3-velocity vectors of the emitter/receiver, namely
\begin{equation}
  (\beta^i)_{A/B}=\left(\frac{u^i}{u^0}\right)_{A/B}=\frac{1}{c}\left(\frac{\dd x^i}{\dd t}\right)_{A/B}\text{.}
  \label{eq:3velAB}
\end{equation}

As shown by \citet{2002PhRvD..66b4045L}, \citet{2012CQGra..29w5027H}, and \citet{2014PhRvD..89f4045H}, the frequency transfer may be expressed in terms of the time (or similarly the range) transfer function after inserting \eqref{eq:kTFdef} and \eqref{eq:4vel} into \eqref{eq:dop}, namely
\begin{equation}
  \frac{\nu_B}{\nu_A}=\frac{(u^0)_B}{(u^0)_A}\frac{q_B}{q_A}\text{,}
  \label{eq:dopTF}
\end{equation}
with
\begin{subequations}\label{eq:qTF}
\begin{align}
  q_A&=1+\beta^i_A\frac{\partial\mathcal{R}}{\partial x_A^i}\text{,}\label{eq:qA}\\
  q_B&=1-\frac{\partial\mathcal{R}}{\partial x_B^0}-\beta^i_B\frac{\partial\mathcal{R}}{\partial x_B^i}\text{.}\label{eq:qB}
\end{align}
\end{subequations}
Therefore, the time and frequency transfers in Eqs.~\eqref{eq:TTFdef} and \eqref{eq:qTF} can be computed once the explicit form of the time (or similarly the range) transfer function is known.

The time transfer function may be determined from the eikonal equation. Indeed, after making use of \eqref{eq:kcovdef} and \eqref{eq:defl} while evaluating \eqref{eq:eik} at $x_A$, we end up with
\begin{equation}
  \left(\bar g^{00}+2\bar g^{0i}l_i+\bar g^{ij}l_il_j\right)_A=0\text{.}
\end{equation}
Then, by invoking Eqs. \eqref{eq:kTFdef}, \eqref{eq:Gorcon}, and assumption \eqref{eq:flat}, we obtain the Hamilton-Jacobi equation that is satisfied by $\mathcal R$. By replacing $\bm x_A$ by a variable $\bm x$ while considering $x_B^0$ and $\bm x_B$ as fixed parameters, the Hamilton-Jacobi equation eventually reads as
\begin{align}
  \big[\partial_i\mathcal R\partial_i\mathcal R\big]_{(\bm x,x_B)}&=1+\kappa^{00}\big(x_B^0-\mathcal R\big(\bm x,x_B\big),\bm x\big)\nonumber\\
  &+2\kappa^{0i}\big(x_B^0-\mathcal R\big(\bm x,x_B\big),\bm x\big)\big[\partial_i\mathcal R\big]_{(\bm x,x_B)}\nonumber\\
  &+\kappa^{ij}\big(x_B^0-\mathcal R\big(\bm x,x_B\big),\bm x\big)\big[\partial_i\mathcal R\partial_j\mathcal R\big]_{(\bm x,x_B)}\text{.}\label{eq:eikdec}
\end{align}

The form of the optical metric in Eqs.~\eqref{eq:Gorcon} and \eqref{eq:Gorcov} implies that the range transfer function can be looked for as
\begin{equation}
  \mathcal{R}(\bm{x},x_B)=\Vert\bm{x}_B-\bm{x}\Vert+\Delta(\bm{x},x_B)\text{,}
  \label{eq:RTFdecR}
\end{equation}
where $\Delta$ is called a refractive delay function. In the present context, the delay function depends on $\kappa^{\mu\nu}$ and is due to refraction when the light ray is crossing through $\mathcal{D}$. After substituting for $\mathcal{R}$ from \eqref{eq:RTFdecR} into \eqref{eq:eikdec}, we find an expression as follows
\begin{equation}
  -2N^i\big[\partial_i\Delta\big]_{(\bm x,x_B)}=W(x,x_B)\text{,}
  \label{eq:part}
\end{equation}
where we introduced
\begin{align}
  W(x,x_B)&=\big(\kappa^{00}-2\kappa^{0i}N^i+\kappa^{ij}N^iN^j\big)_x+2\big(\kappa^{0i}-\kappa^{ij}N^j\big)_x\big[\partial_i\Delta\big]_{(\bm x,x_B)}+\big(\kappa^{ij}-\delta^{ij}\big)_x\big[\partial_i\Delta\partial_j\Delta\big]_{(\bm x,x_B)}\text{,}
\end{align}
with $\bm N$ given by
\begin{equation}
  \bm N=\frac{\bm x_B-\bm x}{\Vert\bm x_B-\bm x\Vert}\text{,}
  \label{eq:Ndir}
\end{equation}
and where the point-event $x$ is defined as
\begin{equation}
  x(\bm x)=\big(x_B^0-\Vert\bm x_B-\bm x\Vert-\Delta(\bm x,x_B),\bm x\big)\text{.}
  \label{eq:xret}
\end{equation}

Since $\bm x$ is a free variable, we choose for convenience to consider the case where $\bm x$ is varying along the straight line segment joining $\bm x_A$ and $\bm x_B$, that is $\bm x=\bm z(\sigma)$ with
\begin{equation}
  \bm z(\sigma)=\bm x_B-\sigma R_{AB}\bm N_{AB}\text{,} \qquad 0\leqslant\sigma\leqslant 1\text{,}
  \label{eq:z-i}
\end{equation}
where $R_{AB}=\Vert\bm x_B-\bm x_A\Vert$. In that respect, we have
\begin{equation}
  \bm N=\bm N_{AB}\text{.}
  \label{eq:NNAB}
\end{equation}
A straightforward calculation shows that the total differentiation of $\Delta(\bm z(\sigma),x_B)$ with respect to $\sigma$ is given by
\begin{equation}
  \frac{\dd}{\dd\sigma}\Delta\big(\bm z(\sigma),x_B\big)=-R_{AB}N_{AB}^i\big[\partial_i\Delta\big]_{(\bm z(\sigma),x_B)}\text{,}
  \label{eq:partot}
\end{equation}
where $[\partial_i\Delta]_{(\bm z(\sigma),x_B)}$ denotes the partial derivative of $\Delta(\bm x,x_B)$ with respect to $x^i$ taken at $\bm x=\bm z(\sigma)$. Then, after inserting \eqref{eq:part} and \eqref{eq:Ndir} into \eqref{eq:partot} while accounting for \eqref{eq:NNAB}, we infer
\begin{equation}
  \frac{\dd}{\dd\sigma}\Delta\big(\bm z(\sigma),x_B\big)=\frac{R_{AB}}{2}W\big(\tilde z(\sigma),x_B\big)\text{,}
  \label{eq:Delfond}
\end{equation}
where the components of the point-event $\tilde z(\sigma)$ are given by \eqref{eq:xret} with $\tilde z(\sigma)\equiv x(\bm z(\sigma))$, namely
\begin{equation}
  \tilde z(\sigma)=\big(x_B^0-\sigma R_{AB}-\Delta(\bm z(\sigma),x_B),\bm z(\sigma)\big)\text{,} \qquad 0\leqslant\sigma\leqslant 1\text{.}
\end{equation}

Equation \eqref{eq:Delfond} is the fundamental differential equation for the determination of the delay function. It can be integrated with the following boundary conditions
\begin{subequations}
\begin{align}
  \Delta\big(\bm z(0),x_B\big)&=0\text{,}\\
  \Delta\big(\bm z(1),x_B\big)&=\Delta\big(\bm x_A,x_B\big)\text{,}
\end{align}
\end{subequations}
which follow from the requirement that $\Delta(\bm x_B,x_B)=0$ and from $\bm z(0)=\bm x_B$. Therefore, \eqref{eq:Delfond} is such that
\begin{align}
  \Delta(\bm x_A,x_B)&=\frac{R_{AB}}{2}\int_{\mathcal D}\bigg\{\big(\kappa^{00}-2\kappa^{0i}N^i_{AB}+\kappa^{ij}N^i_{AB}N^j_{AB}\big)_{\tilde z(\sigma)}\nonumber\\*
    &+2\big(\kappa^{0i}-\kappa^{ij}N_{AB}^j\big)_{\tilde z(\sigma)}\left[\partial_i\Delta\right]_{(\bm{z}(\sigma),x_B)}-\left[\partial_i\Delta\partial_i\Delta\right]_{(\bm{z}(\sigma),x_B)}\bigg\}\,\dd\sigma\text{,}
  \label{eq:Delint}
\end{align}
where the integration goes from $\sigma=0$ to $1$ and is limited to the spacetime region within $\mathcal D$.

Henceforth, we assume that $\Gamma_{AB}$ is almost a straight-line segment so the corresponding time transfer function is unique and admits an analytical expansion. In agreement with this assumption, we suppose that the past null cone at $x_B$, denoted by $\mathscr{C}(x_B)$, intersects $\mathcal{C}_A$ the world-line of the emitter at only one point-event $x_A$ (see Fig.~\ref{fig:lightcone}). Accordingly, \eqref{eq:Delint} can be solved iteratively \citep{PhysRevD.101.064035}.

\subsection{Post-Minkowskian expansion}
\label{sec:detref}

In what follows, we suppose that the refractivity of the medium satisfies the condition
\begin{equation}
  N(x)\ll 1\text{,}
  \label{eq:Ninf1}
\end{equation}
on the domain $\mathcal D$. This condition implies that the optical metric $\bar g$ deviates only slightly from the Minkowski metric, that is to say
\begin{equation}
  |\gamma_{\mu\nu}|\ll 1\text{,} \qquad |\kappa^{\mu\nu}|\ll 1\text{.}\label{eq:approxPM}
\end{equation}
Then, Eqs. \eqref{eq:approxPM} ensure that the null geodesic path $\Gamma_{AB}$ is almost a straight-line segment. 

Let $N_0$ be the refractivity at a well chosen point-event $x_0\in\mathcal{D}$, that is to say $N_0=N(x_0)$. In the context of occultation experiments, it is convenient to consider that $N_0$ is the refractivity at the ground level for a rocky planet or satellite. For gas giants $N_0$ can be defined as the refractivity at a certain pressure level. Thus, for planetary neutral atmospheres, we can naturally assume that
\begin{equation}
  N(x)=N_0\mathcal N(x)\text{,}
  \label{eq:NN0}
\end{equation}
with $ N(x)\leqslant N_0$, where $\mathcal N(x)$ is a function that we assume to be independent of $N_0$ with $0<\mathcal N(x)\leqslant 1$. Therefore, Eq. \eqref{eq:Ninf1} implies that $N_0\ll1$. Thus, in the present context, $N_0$ can be seen as the post-Minkowskian parameter of the theory similarly to what is done with gravity and the gravitational constant $G$. In that sense, any expansion in ascending power of $N_0$ can be qualified as a post-Minkowskian expansion too.

Accordingly, we can always expand the components $\kappa^{\mu\nu}$ in ascending power of $N_0$ such as
\begin{equation}
  \kappa^{\mu\nu}(x,N_0)=\sum_{m=1}^{\infty}(N_0)^m\kappa^{\mu\nu}_{(m)}(x)\text{.}
  \label{eq:kapPM}
\end{equation}
Consequently, the delay function, introduced in Eq. \eqref{eq:RTFdecR}, also admits a post-Minkowskian expansion:
\begin{equation}
  \Delta(\bm{x}_A,x_B,N_0)=\sum_{m=1}^{\infty}(N_0)^m\Delta^{(m)}(\bm{x}_A,x_B)\text{.}
  \label{eq:delRrPMN0G}
\end{equation}
After substituting for $\kappa^{\mu\nu}(\tilde z(\sigma))$ and $\Delta(\bm x_A,x_B)$ from \eqref{eq:kapPM} and \eqref{eq:delRrPMN0G} into \eqref{eq:Delint}, one deduces the expressions for the quantities $\Delta^{(m)}$ as \citep{PhysRevD.101.064035}
\begin{subequations}\label{eq:intdelRrPM}
  \begin{align}
    \Delta^{(1)}(\bm{x}_A,x_B)&=\frac{R_{AB}}{2}\int_{\mathcal D}\left(\kappa^{00}_{(1)}-2\kappa^{0i}_{(1)}N^i_{AB}+\kappa^{ij}_{(1)}N^i_{AB}N^j_{AB}\right)_{z(\sigma)}\dd\sigma\text{,}\label{eq:intdelRrPM1}\\
    \Delta^{(2)}(\bm{x}_A,x_B)&=\frac{R_{AB}}{2}\int_{\mathcal D}\Bigg\{\left(\hat{\kappa}^{00}_{(2)}-2\hat{\kappa}^{0i}_{(2)}N^i_{AB}+\hat{\kappa}^{ij}_{(2)}N^i_{AB}N^j_{AB}\right)_{(z(\sigma),x_B)}\nonumber\\*
    &+2\left(\kappa^{0i}_{(1)}-\kappa^{ij}_{(1)}N_{AB}^j\right)_{z(\sigma)}\left[\frac{\partial\Delta^{(1)}}{\partial x^i}\right]_{(\bm{z}(\sigma),x_B)}-\left[\frac{\partial\Delta^{(1)}}{\partial x^i}\frac{\partial\Delta^{(1)}}{\partial x^i}\right]_{(\bm{z}(\sigma),x_B)}\Bigg\}\dd\sigma\text{,}\label{eq:intdelRrPM2}
  \end{align}
  and, for $l \geqslant 3$ by
  \begin{align}
    \Delta^{(l)}(\bm{x}_A,x_B)&=\frac{R_{AB}}{2}\int_{\mathcal D}\Bigg\{\left(\hat{\kappa}^{00}_{(l)}-2\hat{\kappa}^{0i}_{(l)}N^i_{AB}+\hat{\kappa}^{ij}_{(l)}N^i_{AB}N^j_{AB}\right)_{(z(\sigma),x_B)}\nonumber\\
    &+2\sum_{m=1}^{l-1}\left(\hat\kappa^{0i}_{(m)}-\hat\kappa^{ij}_{(m)}N_{AB}^j\right)_{(z(\sigma),x_B)}\left[\frac{\partial\Delta^{(l-m)}}{\partial x^i}\right]_{(\bm{z}(\sigma),x_B)}-\sum_{m=1}^{l-1}\left[\frac{\partial\Delta^{(m)}}{\partial x^i}\frac{\partial\Delta^{(l-m)}}{\partial x^i}\right]_{(\bm{z}(\sigma),x_B)}\nonumber\\
    &+\sum_{m=1}^{l-2}\left(\hat{\kappa}^{ij}_{(m)}\right)_{(z(\sigma),x_B)}\sum_{n=1}^{l-m-1}\left[\frac{\partial\Delta^{(n)}}{\partial x^i}\frac{\partial\Delta^{(l-m-n)}}{\partial x^j}\right]_{(\bm{z}(\sigma),x_B)}\Bigg\}\dd\sigma\text{.}
  \end{align}
\end{subequations}
The integrations are limited to the refractive domain $\mathcal{D}$ (see Fig.~\ref{fig:lightcone}) and the point-event $z(\sigma)$ is defined by
\begin{equation}
  z(\sigma)=\big(x_B^0-\sigma R_{AB},\bm{z}(\sigma)\big)\text{,} \qquad 0\leqslant\sigma\leqslant 1\text{.}
  \label{eq:z-m}
\end{equation}
The quantities $\hat{\kappa}^{\mu\nu}_{(m)}$ are given by
\begin{subequations}\label{eq:kapRPMdec}
\begin{equation}
  \hat{\kappa}^{\mu\nu}_{(1)}(z(\sigma),x_B)=\kappa^{\mu\nu}_{(1)}(z(\sigma))\text{,}
\end{equation}
and, for $l\geqslant 2$ by
\begin{align}
  \hat{\kappa}^{\mu\nu}_{(l)}(z(\sigma),x_B)&=\kappa^{\mu\nu}_{(l)}(z(\sigma))+\sum_{m=1}^{l-1}\sum_{n=1}^{m}\Phi^{(m,n)}(\bm z(\sigma),x_B)\left[\frac{\partial^n\kappa^{\mu\nu}_{(l-m)}}{(\partial x^0)^n}\right]_{z(\sigma)}\text{.}
\end{align}
\end{subequations}
The reception function $\Phi^{(m,n)}(\bm x,x_B)$ is defined for $m\geqslant 1$ and $1\leqslant n\leqslant m$ \citep{2008CQGra..25n5020T} and is given by
\begin{align}
  \Phi^{(m,n)}(\bm x,x_B)&=\frac{(-1)^n}{n!}\sum_{k_1+\cdots+k_n=m-n}\Bigg[\prod_{i=1}^n\Delta^{(k_i+1)}(\bm x,x_B)\Bigg]\text{,}
  \label{eq:recfun}
\end{align}
with $k_1,\ldots,k_m\in\mathbb{N}_{\geqslant 0}$. The summation in \eqref{eq:recfun} is taken over all sequences of $k_1$ through $k_n$ such that the sum of all $k_n$ is $m-n$.

\section{Application to radio occultations}
\label{sec:app}

The time transfer functions formalism is now applied to static and stationary optical spacetimes. Then, the discussion is specialized to spherical symmetry and to radio occultations considering the case of a downlink one-way transfer.

\subsection{Static optical spacetime}

Let us assume that the optical medium is still in the global coordinate system $(x^\mu)$, namely
\begin{equation}
  w^\mu=(1,\bm 0)\text{.}
\end{equation}
Therefore, the post-Minkowskian expansion of the optical metric can be derived straightforwardly from Eqs. \eqref{eq:Gorcon}. The only non-null components are
\begin{equation}
  \kappa^{00}_{(1)}=2\mathcal N\text{,} \qquad \kappa^{00}_{(2)}=\mathcal N^2\text{.}
  \label{eq:kaptic}
\end{equation}

Within the so-considered coordinate system, the refractive properties of the medium are supposed to be independent of the component $x^0$, that is to say
\begin{equation}
  n(\bm{x})-1=
  \left\{
  \begin{array}{l l}
    N_0\mathcal N(\bm{x}) & \mathrm{for}\ \bm{x}\in\mathcal{D}\text{,}\\
    0 & \mathrm{for}\ \bm x\notin\mathcal{D}\text{.}
  \end{array}
  \right.
\end{equation}
Hence, the optical spacetime is constant and static, so that relations \eqref{eq:kapRPMdec} reduce to
\begin{equation}
  \hat{\kappa}^{\mu\nu}_{(l)}(z(\sigma),x_B)=\kappa^{\mu\nu}_{(l)}(\bm z(\sigma)) \qquad \mathrm{for}\ l\geqslant 1\text{.}\label{eq:kaphat}
\end{equation}

By inserting \eqref{eq:kaptic} into \eqref{eq:intdelRrPM} while considering \eqref{eq:kaphat}, we obtain the integral form of the delay function up to the $l$th post-Minkowskian order
\begin{subequations}\label{eq:deltic}
\begin{align}
  \Delta^{(1)}(\bm{x}_A,\bm x_B)&=R_{AB}\int_{\mathcal D}\big(\mathcal N\big)_{\bm z(\sigma)}\dd\sigma\text{,}\label{eq:deltic1}\\
  \Delta^{(2)}(\bm{x}_A,\bm x_B)&=\frac{R_{AB}}{2}\int_{\mathcal D}\Bigg\{\left(\mathcal N^2\right)_{\bm z(\sigma)}-\left[\frac{\partial\Delta^{(1)}}{\partial x^i}\frac{\partial\Delta^{(1)}}{\partial x^i}\right]_{(\bm z(\sigma),\bm x_B)}\Bigg\}\dd\sigma\text{,}\label{eq:deltic2}
\end{align}
and, for $l\geqslant3$
\begin{equation}
  \Delta^{(l)}(\bm{x}_A,\bm x_B)=-\frac{R_{AB}}{2}\int_{\mathcal D}\ \sum_{m=1}^{l-1}\left[\frac{\partial\Delta^{(m)}}{\partial x^i}\frac{\partial\Delta^{(l-m)}}{\partial x^i}\right]_{(\bm{z}(\sigma),\bm x_B)}\dd\sigma\text{.}\label{eq:delticl}
\end{equation}
\end{subequations}

As mentioned previously by \citet{PhysRevD.101.064035}, these equations show that the first order delay is the well-known excess path delay due to the change of the phase velocity when the signal is crossing through $\mathcal D$. The geometric delay shows up at the second post-Minkowskian order as well as the second order correction to the excess path delay.

\subsection{Stationary optical spacetime}

Let us now assume that the fluid optical medium is at rest in a coordinate system rotating with respect to the global coordinate system. Let $(x^{\hat{\alpha}})$ denote the rotating coordinate system \footnote{In this section, we use the convention that hated index starting from the first part of the Greek or Latin alphabet denote components expressed in the rotating frame.}. The relation between the optical medium rest frame and the global coordinate system reads as follows
\begin{subequations}\label{eq:trans}
\begin{equation}
  x^{\hat 0}=x^0\text{,} \qquad x^{\hat a}=\Lambda^{\hat a}_{\ i}(t)x^i\text{,}
  \label{eq:2rot}
\end{equation}
in which $\Lambda^{\hat a}_{\ i}(t)\in\mathrm{SO}(3)$ are the elements of a rotation matrix. The inverse relation is given by
\begin{equation}
  x^0=x^{\hat  0}\text{,} \qquad x^i=\Lambda^i_{\ \hat a}(t)x^{\hat a}\text{,}
  \label{eq:2ine}
\end{equation}
\end{subequations}
with $\Lambda^i_{\ \hat a}(t)$ being the elements of the inverse matrix. The following relationships
\begin{equation}
  \Lambda^{\hat a}_{\ i}\Lambda^i_{\ \hat b}=\delta^{\hat a}_{\ \hat b}\text{,} \qquad \Lambda^i_{\ \hat a}\Lambda^{\hat a}_{\ j}=\delta^{i}_{\ j}\text{,}
\end{equation}
ensure that a transformation followed by its inverse is an identity transformation.

Differentiating Eqs. \eqref{eq:2ine} with respect to the global coordinate time returns the transformation law that relates the 3-velocities expressed in the two coordinate systems
\begin{equation}
  \dot x^{i}=\Lambda^{i}_{\ \hat a}(\dot x^{\hat a}+\omega^{\hat a}_{\ \hat b}x^{\hat b})\text{.}
  \label{eq:dxine}
\end{equation}
An overdot indicates differentiation with respect to $t$ and $\omega_{\hat a\hat c}$ is the angular velocity tensor being defined by
\begin{equation}
  \omega^{\hat a}_{\ \hat c}=\Lambda^{\hat a}_{\ i}\dot\Lambda^{i}_{\ \hat c}\text{,} \qquad \omega_{\hat a\hat c}=\varepsilon_{\hat a\hat b\hat c}\omega^{\hat b}\text{.}
  \label{eq:omtens}
\end{equation}
The quantities $\varepsilon_{\hat a\hat b\hat c}$ represent the components of the permutation symbol and $\omega^{\hat a}$ is the $\hat a$th component of the angular velocity vector expressed in the rotating system, namely
\begin{equation}
  \omega^{\hat a}=\omega e^{\hat a}\text{,}
  \label{eq:omvec}
\end{equation}
where $\omega$ is the magnitude of the angular velocity of rotation and $e^{\hat a}$ is the $\hat a$th component of the unit direction of the spin axis.

Hereafter, we assume that the time dependent rotation matrix corresponds to a rigid uniform rotation so that the angular velocity tensor is constant, namely
\begin{equation}
  \dot\omega_{\hat a\hat c}=0\text{.}
  \label{eq:orot0}
\end{equation}
In addition, in the rotating frame, which is the rest frame of the medium, the 3-velocity vector of a fluid's particle is null by definition, that is to say
\begin{equation}
  \dot x^{\hat a}=0\text{.}
  \label{eq:vrot0}
\end{equation}
The expression for the velocity of a fluid's particle expressed in the global coordinate system is derived after substituting for $\dot x^{\hat a}$ from \eqref{eq:vrot0} into \eqref{eq:dxine} which leads to
\begin{equation}
  \frac{\dd x^i}{\dd t}=\Lambda^i_{\ \hat a}\omega^{\hat a}_{\ \hat b}\Lambda^{\hat b}_{\ j}x^j\text{.}
  \label{eq:dxidt}
\end{equation}

Let $\bm\xi(x)$ be a coordinate 3-velocity vector field at the point-event $x$ belonging to a fluid element of the optical medium. It is defined in global coordinate notation by
\begin{equation}
  \xi^i(x)=\frac{w^i}{w^0}=\frac{1}{c}\frac{\dd x^i}{\dd t}\text{.}
  \label{eq:3velel}
\end{equation}
Owing to \eqref{eq:omtens}, \eqref{eq:omvec}, and \eqref{eq:dxidt}, the coordinate 3-velocity vector of an element of the fluid dielectric medium is given in the global coordinate system by
\begin{equation}
  \bm\xi(\bm x)=\frac{\omega}{c}\bm e\times\bm x\text{.}
  \label{eq:3velmed}
\end{equation}
Thus, the unit 4-velocity vector of the medium reads as follows
\begin{equation}
  w^{\mu}(\bm x)=w^0\big(1,\bm \xi(\bm x)\big)\text{.}
  \label{eq:4velmed}
\end{equation}
The expression for the time component $w^0$ is straightforwardly inferred from the fact that the 4-velocity is a unit vector for the physical metric of spacetime (see assumption \eqref{eq:flat}), hence $w^0=\Gamma$, where $\Gamma$ is defined as
\begin{equation}
  \Gamma(\bm x)=\frac{1}{\sqrt{1-\bm\xi(\bm x)\cdot\bm\xi(\bm x)}}\text{.}
  \label{eq:u0}
\end{equation}
We can now express the components of the optical metric from the relations \eqref{eq:Gorcon}. The only non-null components are
\begin{subequations}\label{eq:kapGCRS}
\begin{align}
  \kappa^{00}_{(1)}&=2\Gamma^2\mathcal N\text{,} \qquad \kappa^{0i}_{(1)}=\kappa^{00}_{(1)}\,\xi^i\text{,} \qquad \kappa^{ij}_{(1)}=\kappa^{00}_{(1)}\,\xi^i\xi^j\text{,}\label{eq:kapGCRS1}\\
  \kappa^{00}_{(2)}&=\Gamma^2\mathcal N^2\text{,} \qquad \kappa^{0i}_{(2)}=\kappa^{00}_{(2)}\,\xi^i\text{,} \qquad \kappa^{ij}_{(2)}=\kappa^{00}_{(2)}\,\xi^i\xi^j\text{.}\label{eq:kapGCRS2}
\end{align}
\end{subequations}

As seen in Sec. \ref{sec:relopt}, the index of refraction is defined in the instantaneous rest frame of the medium namely the rotating frame. Thus, the index of refraction is independent of the time component of the rotating coordinate system, that is to say
\begin{equation}
  n(x^{\hat a})-1=
  \left\{
  \begin{array}{l l}
    N_0\mathcal N(x^{\hat a}) & \mathrm{for}\ x^{\hat a}\in\mathcal{D}\text{,}\\
    0 & \mathrm{for}\ x^{\hat a}\notin\mathcal{D}\text{.}
  \end{array}
  \right.
\end{equation}
This statement implies that $\partial_0\mathcal N=0$ according to the transformation rules in \eqref{eq:trans}. Similarly, \eqref{eq:3velmed} also reveals that $\partial_0\xi^i=0$, hence $\partial_0\Gamma=0$. These simplifications imply that the optical spacetime is stationary, so Eqs. \eqref{eq:kapRPMdec} eventually reduce to
\begin{equation}
  \hat{\kappa}^{\mu\nu}_{(l)}(z(\sigma),x_B)=\kappa^{\mu\nu}_{(l)}(\bm z(\sigma)) \qquad \mathrm{for}\ l\geqslant 1\text{.}
  \label{eq:kaphat2}
\end{equation}

By inserting \eqref{eq:kapGCRS} into \eqref{eq:intdelRrPM} while considering \eqref{eq:kaphat2}, we obtain the integral form of the delay function up to the $l$th post-Minkowskian order
\begin{subequations}\label{eq:delPM}
\begin{align}
  \Delta^{(1)}(\bm{x}_A,\bm x_B)&=R_{AB}\int_{\mathcal D}\left(\Gamma^2\mathcal N C^2\right)_{\bm z(\sigma)}\dd\sigma\text{,}\label{eq:delPM1}\\
  \Delta^{(2)}(\bm{x}_A,\bm x_B)&=\frac{R_{AB}}{2}\int_{\mathcal D}\Bigg\{\left(\Gamma^2\mathcal N^2C^2\right)_{\bm z(\sigma)}-\left[\frac{\partial\Delta^{(1)}}{\partial x^i}\frac{\partial\Delta^{(1)}}{\partial x^i}\right]_{(\bm z(\sigma),\bm x_B)}+4\left(\Gamma^2\mathcal NC\xi^i\right)_{\bm z(\sigma)}\left[\frac{\partial\Delta^{(1)}}{\partial x^i}\right]_{(\bm z(\sigma),\bm x_B)}\Bigg\}\dd\sigma\text{,}\label{eq:delPM2}\\
  \Delta^{(3)}(\bm{x}_A,\bm x_B)&=R_{AB}\int_{\mathcal D}\Bigg\{\left(\Gamma^2\mathcal N\xi^i\xi^j\right)_{\bm z(\sigma)}\left[\frac{\partial\Delta^{(1)}}{\partial x^i}\frac{\partial\Delta^{(1)}}{\partial x^j}\right]_{(\bm z(\sigma),\bm x_B)}+\left(\Gamma^2\mathcal N^2C\xi^i\right)_{\bm z(\sigma)}\left[\frac{\partial\Delta^{(1)}}{\partial x^i}\right]_{(\bm z(\sigma),\bm x_B)}\nonumber\\
  &+2\left(\Gamma^2\mathcal NC\xi^i\right)_{\bm z(\sigma)}\left[\frac{\partial\Delta^{(2)}}{\partial x^i}\right]_{(\bm z(\sigma),\bm x_B)}-\left[\frac{\partial\Delta^{(1)}}{\partial x^i}\frac{\partial\Delta^{(2)}}{\partial x^i}\right]_{(\bm z(\sigma),\bm x_B)}\Bigg\}\dd\sigma\text{,}
\end{align}
and, for $l\geqslant 4$
\begin{align}
  \Delta^{(l)}(\bm{x}_A,\bm x_B)&=\frac{R_{AB}}{2}\int_{\mathcal D}\Bigg\{4\left(\Gamma^2\mathcal NC\xi^i\right)_{\bm z(\sigma)}\left[\frac{\partial\Delta^{(l-1)}}{\partial x^i}\right]_{(\bm z(\sigma),\bm x_B)}+2\left(\Gamma^2\mathcal N^2C\xi^i\right)_{\bm z(\sigma)}\left[\frac{\partial\Delta^{(l-2)}}{\partial x^i}\right]_{(\bm z(\sigma),\bm x_B)}\nonumber\\
  &-\sum_{m=1}^{l-1}\left[\frac{\partial\Delta^{(m)}}{\partial x^i}\frac{\partial\Delta^{(l-m)}}{\partial x^i}\right]_{(\bm z(\sigma),\bm x_B)}+2\left(\Gamma^2\mathcal N\xi^i\xi^j\right)_{\bm z(\sigma)}\sum_{n=1}^{l-2}\left[\frac{\partial\Delta^{(n)}}{\partial x^i}\frac{\partial\Delta^{(l-n-1)}}{\partial x^j}\right]_{(\bm z(\sigma),\bm x_B)}\nonumber\\
  &+\left(\Gamma^2\mathcal N^2\xi^i\xi^j\right)_{\bm z(\sigma)}\sum_{n=1}^{l-3}\left[\frac{\partial\Delta^{(n)}}{\partial x^i}\frac{\partial\Delta^{(l-n-2)}}{\partial x^j}\right]_{(\bm z(\sigma),\bm x_B)}\Bigg\}\dd\sigma\text{.}
\end{align}
\end{subequations}
We introduced $C$ and $D$ as
\begin{equation}
  C(\bm x)=1-D(\bm x)\text{,} \qquad D(\bm x)=\bm \xi(\bm x)\cdot\bm N_{AB}\text{.}
  \label{eq:CD}
\end{equation}
Hereafter, $C$ is referred to as the geometric factor and $D$ as the light-dragging term.

By comparing \eqref{eq:delPM1} and \eqref{eq:deltic1}, it is seen that the dynamics of the optical medium affects the expression of the delay function as soon as the first post-Minkowskian order. Hereafter, we solve \eqref{eq:delPM1} assuming a spherically symmetric optical spacetime.

\subsection{Spherical symmetry}
\label{sec:sphesym}

Let us assume that the optical medium is the planetary neutral atmosphere of the occulting body. The global coordinate system is supposed to be centered at the center of mass of the occulting body and is non-rotating with respect to distant stars. The medium is assumed to be at rest in the frame rotating with the planet (i.e. the medium rest frame). The atmosphere of the occulting body is assumed to be spherically symmetric so centrifugal effects due to the rotation are neglected. In that respect, $\mathcal{D}$ draws a timelike tube defining the spacetime boundaries of the planetary neutral atmosphere (see Fig.~\ref{fig:lightcone}).

The spherical symmetry assumption allows us to determine uniquely the limits of integration in \eqref{eq:delPM}. Let $\mathcal{H}$ be the radii of the top neutral atmosphere. The intersection between the path of integration and $\mathcal H$ can be determined from \eqref{eq:z-i} as
\begin{equation}
 \Vert\bm z(\sigma)\Vert=\mathcal{H}\text{.}
\end{equation}
After little algebra, we find
\begin{equation}
  \sigma_{\pm}=\sigma_K\pm\frac{\sqrt{\mathcal H^2-K^2}}{R_{AB}}\text{,} \qquad \sigma_K=\frac{\bm N_{AB}\cdot\bm x_B}{R_{AB}}\text{,}
  \label{eq:lpmK}
\end{equation}
where we have introduced
\begin{equation}
  K=\Vert\bm N_{AB}\times\bm x_B\Vert\text{.}
  \label{eq:K}
\end{equation}
This last expression suggests that $K$ is the impact parameter with respect to the center of symmetry (i.e. the center of mass of the occulting planet).

Let $x_K$ be the point-event defined by $x_K\equiv z(\sigma_{K})$. After substituting for $\sigma_K$ from \eqref{eq:lpmK} into \eqref{eq:z-m}, we find the spacetime components of $x_K=(x_K^0,\bm x_K)$ with
\begin{subequations}\label{eq:peK}
\begin{align}
  x_K^0&=x_B^0-\bm N_{AB}\cdot\bm x_B\text{,}\label{eq:ctK}\\
  \bm x_K&=(\bm N_{AB}\times\bm x_B)\times\bm N_{AB}\text{.}\label{eq:xK}
\end{align}
\end{subequations}
It can be seen that $K=\Vert\bm x_K\Vert$, thus the unit 3-vector for the direction of $\bm x_K$, namely $\bm n_K=\bm x_K/K$, is given by
\begin{equation}
  \bm n_K=\frac{\bm N_{AB}\times\bm x_B}{\Vert\bm N_{AB}\times\bm x_B\Vert}\times\bm N_{AB}\text{.}
  \label{eq:nK}
\end{equation}
Therefore, $x_K$ is the point-event along the path of integration where the euclidean distance with respect to the center of symmetry is the smallest. Let us notice that \eqref{eq:nK} implies that
\begin{equation}
  \bm n_K\cdot\bm N_{AB}=0\text{.}
  \label{eq:nKNAB}
\end{equation}
Accordingly, it is helpful to introduce the unit 3-vector $\bm S_{AB}$ such that the triad of vectors $(\bm n_K,\bm N_{AB},\bm S_{AB})$ forms a right-handed vector basis, that is to say
\begin{equation}
  \bm S_{AB}=\bm n_K\times\bm N_{AB}\text{.}
  \label{eq:nKxNAB}
\end{equation}
After identifying \eqref{eq:nKxNAB} with \eqref{eq:nK}, we deduce
\begin{equation}
  \bm S_{AB}=-\frac{\bm N_{AB}\times\bm x_B}{\Vert\bm N_{AB}\times\bm x_B\Vert}\text{.}
  \label{eq:sig}
\end{equation}
The unit vector $\bm S_{AB}$ is thus recognized to be the direction of the angular momentum vector of the zeroth-order null geodesic path \citep{2012CQGra..29x5010T}. 

Let $x_+$ and $x_-$ be the point-events defined by $x_{\pm}\equiv z(\sigma_{\pm})$. After substituting for $\sigma_{\pm}$ from \eqref{eq:lpmK} into \eqref{eq:z-m}, we find the spacetime components of $x_{\pm}=(x^0_{\pm},\bm{x}_{\pm})$ with
\begin{subequations}\label{eq:pe+-}
\begin{align}
  x^0_{\pm}&=x^0_K\mp\sqrt{\mathcal H^2-K^2}\text{,}\label{eq:ct+-}\\
  \bm x_{\pm}&=\bm x_K\mp\sqrt{\mathcal H^2-K^2}\bm N_{AB}\text{.}\label{eq:x+-}
\end{align}
\end{subequations}
The point-event $x_+$ is the spacetime point where the zeroth-order null geodesic path is entering $\mathcal D$, and conversely, $x_-$ is the spacetime point where the zeroth-order null geodesic path is exiting $\mathcal D$ as shown in Fig.~\ref{fig:lightcone}.

Therefore, in the context of radio occultations by a spherically symmetric atmosphere, any integral over the refractive domain, such as
\begin{equation}
  I(\bm x_A,\bm x_B)=\frac{R_{AB}}{2}\int_{\mathcal D}f\big(\bm z(\sigma)\big)\,\dd\sigma\text{,}
  \label{eq:intdef}
\end{equation}
where $f$ is a known function which varies over the path of integration, can now be written as
\begin{equation}
  I(\bm x_A,\bm x_B)=\frac{R_{AB}}{2}\left[\int_{\sigma_-}^{\sigma_K}f\big(\bm z(\sigma)\big)\,\dd\sigma+\int_{\sigma_K}^{\sigma_+}f\big(\bm z(\sigma)\big)\,\dd\sigma\right]\text{.}
  \label{eq:int}
\end{equation}
By separating the path of integration as
\begin{subequations}\label{eq:zeta+-}
\begin{align}
  \bm y_+(\chi)&=\bm x_K-\chi\sqrt{\mathcal H^2-K^2}\bm N_{AB}\text{,}\\
  \bm y_-(\chi)&=\bm x_--\chi\sqrt{\mathcal H^2-K^2}\bm N_{AB}\text{,}
\end{align}
with
\begin{equation}
  0\leqslant\chi\leqslant 1\text{,}
  \label{eq:conmu}
\end{equation}
\end{subequations}
then, by making use of \eqref{eq:xK}, \eqref{eq:x+-}, and \eqref{eq:z-i}, one infers that \eqref{eq:int} can also be written as
\begin{align}
  I(\bm x_A,\bm x_B)&=\frac{\sqrt{\mathcal H^2-K^2}}{2}\left[\int_{0}^{1}f\big(\bm y_-(\chi)\big)\,\dd\chi+\int_{0}^{1}f\big(\bm y_+(\chi)\big)\,\dd\chi\right]\text{.}\label{eq:int1}
\end{align}

The spherical symmetry implies that $f=f(r)$ so it is more convenient to integrate toward the radial component $r$. To perform the change of variable, we can introduce $r$ as follows
\begin{equation}
  r=\Vert\bm y_+(\chi)\Vert\text{,} \qquad r=\Vert\bm y_-(\chi)\Vert\text{,}
\end{equation}
and resolve for $\chi$ considering condition \eqref{eq:conmu}, that is to say
\begin{equation}
  \chi_+(r)=\frac{\sqrt{r^2-K^2}}{\sqrt{\mathcal H^2-K^2}}\text{,} \qquad \chi_-(r)=1-\chi_+(r)\text{.}
\end{equation}
The expressions for the differentials of $\chi_\pm$ are given by
\begin{equation}
  \dd\chi_\pm=\frac{\pm r\,\dd r}{\sqrt{\mathcal H^2-K^2}\sqrt{r^2-K^2}}\text{,}
\end{equation}
such that \eqref{eq:int1} now reads
\begin{subequations}
\begin{equation}
  I(K,\mathcal H)=\int_K^{\mathcal{H}}f\big(\bm y(r)\big)\frac{r\,\dd r}{\sqrt{r^2-K^2}}\text{,}
  \label{eq:int2}
\end{equation}
where the path of integration satisfies $\Vert\bm y(r)\Vert=r$ and is given by
\begin{equation}
  \bm y(r)=\bm x_K-\sqrt{r^2-K^2}\bm N_{AB}\text{.}
  \label{eq:y+r}
\end{equation}
\end{subequations}

Because relations \eqref{eq:delPM} are the same than \eqref{eq:intdef}, which is itself equivalent to \eqref{eq:int2}, the 3-velocity of the medium must be evaluated along $\bm y(r)$ within the assumption of spherical symmetry. From \eqref{eq:y+r} and \eqref{eq:3velmed}, we infer
\begin{equation}
  \bm\xi\big(\bm y(r)\big)=\bm\Omega\times\left(\bm n_K-\frac{\sqrt{r^2-K^2}}{K}\bm N_{AB}\right)\text{,}
  \label{eq:xiOme}
\end{equation}
where we introduced
\begin{equation}
  \bm\Omega=\Omega\bm e\text{,} \qquad \Omega=\frac{\omega K}{c}\text{.}
\end{equation}

We can immediately see from \eqref{eq:nKNAB}, \eqref{eq:nKxNAB}, and \eqref{eq:xiOme}, that the scalar product of the dragging term in \eqref{eq:CD} is actually independent of $r$ and can therefore be considered constant during integration along $\bm y(r)$. Thus, the light-dragging coefficient reads
\begin{equation}
  D=\bm\Omega\cdot\bm S_{AB}\text{.}
  \label{eq:D}
\end{equation}
According to \eqref{eq:CD}, we deduce that the geometric factor $C$ is independent of $r$ too.

\section{Mathematical modeling}
\label{sec:mathmod}

According to relations \eqref{eq:delPM}, we now need a mathematical expression describing the radial evolution of refractivity in order to derive the expressions for the time/frequency transfers.

We emphasize that the method usually employed for processing radio occultation data proceeds the other way around. Indeed, the refractivity profile is usually determined from the frequency transfer by employing Abel inversion \citep{1968JGR....73.1819P} or numerical ray-tracing \citep{2015RaSc...50..712S} methods. Here, the approach is more closely related to a model fitting parameter method. Indeed, we first build a mathematical modeling for the refractivity profile and then we deduce the consequences at the level of the observables, namely the time/frequency transfers. In principle, the last step should be to compare these computed observables to real ones in order to minimize the differences by estimating the parameters of the model (i.e. the parameters entering the refractivity profile) using e.g. a standard least-squares fit.

In appendix \ref{sec:Abel}, we comment about how the ideas of Sec. \ref{sec:app} can indeed be applied in the context of an Abel inversion method while accounting for the light-dragging effect, such that no \emph{a priori} modeling for the refractive profile is required.

\subsection{Refractivity profile}
\label{sec:Nh}

In the context of atmospheric occultation experiments an \emph{a priori} knowledge of the atmospheric composition must be assumed. From the composition one can determine the refractive volume $N_v$ and then the mean density which eventually leads to the refractivity by making use of the ideal gas law:
\begin{equation}
  P(h)=\Bigg(\frac{N}{N_v}\Bigg)_h kT(h)\text{.}
  \label{eq:PGL}
\end{equation}
In this expression, $P$ is the pressure profile, $T$ is the temperature profile, and $k$ is the Boltzmann constant, namely
\begin{equation}
  k=1.380\,649\times10^{-23}\ \mathrm{m}^2\cdot\mathrm{kg}\cdot\mathrm{s}^{-2}\cdot\mathrm{K}^{-1}\text{.}
\end{equation}
The altitude above the ground level $R$ (i.e. the altitude at $\bm x_0$ where $N(\bm x_0)=N_0$, see discussion in Sec. \ref{sec:detref}) is denoted by $h$ and is given by
\begin{equation}
  h=r-R\text{.}
  \label{eq:defh}
\end{equation}
From Eq. \eqref{eq:PGL}, we deduce an expression as follows
\begin{equation}
  N(h)=N_0\,\Bigg(\frac{P}{P_0}\Bigg)_h\Bigg(\frac{T_0}{T}\Bigg)_h\text{,}
  \label{eq:Nh}
\end{equation}
where $P_0$ and $T_0$ are the pressure and temperature at the ground level, respectively. The refractivity at the ground level $N_0$ is given by
\begin{equation}
  N_0=N_v\,\frac{P_0}{kT_0}\text{.}
\end{equation}

The refractivity profile in \eqref{eq:Nh} is eventually proportional to the product of two functions, namely $(P/P_0)_h$ and $(T_0/T)_h$. However, let us emphasis that for planetary atmospheres the temperature usually varies much more slowly than the pressure across the profile. Therefore, in some application that does not necessitate high precision, it might be convenient to consider that $(T_0/T)_h$ is constant (isotherm atmosphere) with respect to $(P/P_0)_h$. In this work, we do not make such a simplification and we consider that the temperature is a function of the altitude inside the atmosphere.

Planetary neutral atmospheres all admit an exponential pressure profile as a first approximation so that it is common to model the pressure as
\begin{equation}
  \Bigg(\frac{P}{P_0}\Bigg)_h=\exp\left(-\frac{h}{H}\right)\text{,}
  \label{eq:ph}
\end{equation}
where $H$ is a constant parameter called the scale height of the neutral atmosphere and has length dimension ($\mathrm{L}$). Then, the large scale temperature variation across the atmospheric profile can be expressed as a polynomial function of degree $d$, namely
\begin{equation}
  \Bigg(\frac{T_0}{T}\Bigg)_h=\sum_{m=0}^{d}a_mh^m\text{,}
  \label{eq:T0sTh}
\end{equation}
where $a_m$ are the polynomial coefficients and have dimension $\mathrm{L}^{-m}$. Because $T_0$ is the temperature at the ground level (i.e. $h=0\ \mathrm{km}$), we have a relation as follows
\begin{equation}
  a_0=1\text{.}
\end{equation}

The series expansion in \eqref{eq:T0sTh} is easily changed into a function of $r$ with \eqref{eq:defh}. After little algebra, we find
\begin{equation}
  \Bigg(\frac{T_0}{T}\Bigg)_r=\sum_{m=0}^{d}b_mr^m\text{,}
  \label{eq:T0sT}
\end{equation}
where the $b_m$ coefficients have dimension $\mathrm{L}^{-m}$ and are given by
\begin{equation}
  b_m=\frac{1}{R^m}\sum_{l=m}^{d}\left(
  \begin{array}{c}
    l\\
    m
  \end{array}
  \right)(-1)^{l-m}a_lR^{l}\text{.}
  \label{eq:bm}
\end{equation}
The binomial coefficient is defined as
\begin{equation}
  \left(
  \begin{array}{c}
    l\\
    m
  \end{array}
  \right)=\frac{l!}{m!(l-m)!}\text{.}
\end{equation}

In the context of radio occultation experiments, it is more convenient to express the profiles in term of $\eta$ the altitude above the impact parameter, namely
\begin{equation}
  \eta=r-K\text{.}
  \label{eq:reta}
\end{equation}
The advantage for using $\eta$ instead of $r$ relies on the fact that most of applications satisfy $\eta\ll K$ everywhere in $\mathcal D$ which allows us to look for solution as infinite series in ascending power of $\eta/K$. Accordingly, the temperature variation now reads
\begin{equation}
  \Bigg(\frac{T_0}{T}\Bigg)_{\eta}=\sum_{m=0}^{d}B_m\left(\frac{\eta}{K}\right)^m\text{,}
  \label{eq:T0sTeta}
\end{equation}
with
\begin{subequations}\label{eq:Bm}
\begin{equation}
  B_m(K)=\sum_{l=m}^{d}\left(
  \begin{array}{c}
    l\\
    m
  \end{array}
  \right)b_lK^l\text{,}\label{eq:Bmb}\\
\end{equation}
or, after substituting for $b_l$ from \eqref{eq:bm}
\begin{equation}
  B_m(K)=\sum_{l=m}^d\left(
  \begin{array}{c}
    l\\
    m
  \end{array}
  \right)\left(\frac{K}{R}\right)^l\sum_{k=l}^{d}\left(
  \begin{array}{c}
    k\\
    l
  \end{array}
  \right)(-1)^{k-l}a_kR^k\text{.}\label{eq:Bma}
\end{equation}
\end{subequations}
Once expressed in term of $\eta$, the pressure profile now reads
\begin{equation}
  \Bigg(\frac{P}{P_0}\Bigg)_{\eta}=\Bigg(\frac{P_K}{P_0}\Bigg)\exp\left(-\frac{\eta}{H}\right)\text{,}
  \label{eq:peta}
\end{equation}
where $P_K$ is the pressure at the level of the impact parameter and is given by \eqref{eq:ph}, that is to say
\begin{equation}
  \Bigg(\frac{P_K}{P_0}\Bigg)=\exp\left(-\frac{K-R}{H}\right)\text{.}
\end{equation}

The expression for $\mathcal N(\eta)$ can be inferred after inserting \eqref{eq:peta} and \eqref{eq:T0sTeta} into \eqref{eq:Nh}. Finally, by invoking the definition \eqref{eq:NN0}, we eventually find
\begin{equation}
  \mathcal N(\eta)=\left(\frac{P_K}{P_0}\right)\exp\left(-\frac{\eta}{H}\right)\sum_{m=0}^{d}\left(\frac{\eta}{K}\right)^mB_m(K)\text{.}
  \label{eq:NetaSimp}
\end{equation}

Let us evaluate this last relationship at the top of the atmosphere and for the optical ray grazing event, namely $\eta=\mathcal H-K$ with $K=\mathcal H$. A simple substitution into \eqref{eq:NetaSimp} returns
\begin{equation}
  \mathcal N_{\mathcal H}=\left(\frac{P_{\mathcal H}}{P_0}\right)B_0(\mathcal H)\text{.}
  \label{eq:NH}
\end{equation}
This result shows that refractivity is non-null on the limits of the refractive domain $\mathcal D$. In order to ensure a smooth transition between the inside of the domain $\mathcal D$ (where the refractivity should be $N\neq 0$) and the outside (where the refractivity should be $N=0$), we would rather introduce $\mathcal N(\eta)$ such as
\begin{equation}
  \mathcal N(\eta)=\left(\frac{P_K}{P_0}\right)\exp\left(-\frac{\eta}{H}\right)\sum_{m=0}^{d}\left(\frac{\eta}{K}\right)^mB_m(K)-\mathcal N_{\mathcal H}\text{.}
  \label{eq:Neta}
\end{equation}

In the context of radio occultation experiments, the value of $\mathcal H$ should be adjusted such that the effect of $\mathcal N_{\mathcal H}$ becomes unobservable, that is to say $\mathcal N_{\mathcal H}=0$. As seen from \eqref{eq:NH}, this can be achieved by taking the limit $\mathcal H\to\infty$. Hereafter, we continue the discussion keeping $\mathcal H$ to an arbitrary value for completeness.

In practice, parameters $H$ and $b_m$ (or $a_m$) would now need to be determined by confrontation with observations. To do so, we need to derive the expressions for the time and the frequency transfers resulting from \eqref{eq:Neta}.

\subsection{The time transfer function}
\label{sec:TTFRO}

A direct integration of relations \eqref{eq:delPM} is difficult in the context of a purely post-Minkowskian expansion in ascending power of $N_0$. The main difficulty is related to the arbitrariness in the magnitude of the velocity of the medium. However, everywhere within the Solar system, we are only dealing with planetary atmospheres with $\Omega\ll 1$. Consequently, $\Gamma^2$ in Eq. \eqref{eq:u0}:
\begin{equation}
  \Gamma^2(r)=\Big[1-\bm\xi\big(\bm y(r)\big)\cdot\bm\xi\big(\bm y(r)\big)\Big]^{-1}\text{,}
  \label{eq:gam2}
\end{equation}
can be expanded as
\begin{align}
  \Gamma^2(\eta)&=1+\sum_{m=1}^{\infty}\Omega^{2m}\sum_{i+j+k=m}\sum_{p=0}^{\frac{j}{2}+k}2^{j+p}\left(
  \begin{array}{c}
    m\\
    i,j,k
  \end{array}
  \right)\left(
  \begin{array}{c}
    \frac{j}{2}+k\\
    p
  \end{array}\right)\left(\frac{\eta}{K}\right)^{j+2k-p}\times\nonumber\\
  &\times\sum_{q=0}^{2i}\sum_{l=0}^{2k}\left(
  \begin{array}{c}
    2i\\
    q
  \end{array}
  \right)\left(
  \begin{array}{c}
    2k\\
    l
  \end{array}
  \right)(-1)^{q+l}(\bm e\cdot\bm n_{K})^{2q+j}(\bm e\cdot\bm N_{AB})^{2l+j}\text{,}
\end{align}
where the multinomial coefficient is defined by
\begin{equation}
  \left(
  \begin{array}{c}
    k\\
    n_1,n_2,\ldots,n_m
  \end{array}
  \right)=\frac{k!}{n_1!n_2!\cdots n_m!}\text{.}
\end{equation}
This expansion shows that the first non-constant contribution in $\Gamma^2$ is a second order term in $\Omega$.

Hereafter, in order to simplify computations, we consider terms up to first order in $\Omega$, therefore we now assume
\begin{equation}
  \Gamma^2=1+O(\Omega^{2})\text{.}
\end{equation}
The geometric factor (introduced in Eq. \eqref{eq:CD}) is the only term that is contributing at first order in $\Omega$
\begin{equation}
  C^2=1-2\bm\Omega\cdot\bm S_{AB}+O(\Omega^{2})\text{.}
  \label{eq:C2}
\end{equation}

By making use of \eqref{eq:intdef} and \eqref{eq:int2}, we see that \eqref{eq:delPM1} can be written as follows
\begin{equation}
  \Delta^{(1)}(K,\mathcal H)=2C^2\int_K^{\mathcal{H}}\big(\mathcal{N}\big)_{\bm y(r)}\frac{r\,\dd r}{\sqrt{r^2-K^2}}+O(\Omega^{2})\text{.}
  \label{eq:del1KHdef}
\end{equation}
Then, by invoking \eqref{eq:reta}, the function to be integrated can be written as an expansion in ascending power of $\eta/K$, that is to say
\begin{align}
  \Delta^{(1)}(K,\mathcal H)&=C^2\sqrt{2K}\,\sum_{m=0}^{\infty}Q_m\int_{0}^{\mathcal{H}-K}\frac{\dd\eta}{\sqrt{\eta}}\left(\frac{\eta}{K}\right)^m\big(\mathcal{N}\big)_{\bm y(\eta)}+O(\Omega^{2})\text{,}\label{eq:del1KH}
\end{align}
where $Q_m$ is given by
\begin{equation}
  Q_m=\frac{(-1)^{m+1}}{m!}\frac{(2m+1)\cdot(2m-3)!!}{2^{2m}}\text{.}
\end{equation}
The double factorial \citep{garfken67math} is defined by 
\begin{subequations}
\begin{equation}
  m!!=\left\{
  \begin{array}{l l}
    m\times(m-2)\times\ldots\times 3\times 1 & \mathrm{for}\ m\ \mathrm{odd}\text{,}\\
    m\times(m-2)\times\ldots\times 4\times 2 & \mathrm{for}\ m\ \mathrm{even}\text{,}\\
    1 & \mathrm{for}\ m=-1,0\text{,}
  \end{array}
  \right.
\end{equation}
and by
\begin{equation}
  (-2m-1)!!=\frac{(-1)^m}{(2m-1)!!} \qquad \mathrm{for}\ m\geqslant 1\text{.}
\end{equation}
\end{subequations}

After substituting for $\mathcal N(\eta)$ from \eqref{eq:Neta} into \eqref{eq:del1KH}, we arrive to the following expression
\begin{align}
  \Delta^{(1)}(K,\mathcal H)&=C^2\,\Bigg[\sum_{m=0}^{\infty}\mathcal{L}_m(K,\mathcal{H})\sum_{n=0}^{m_d}Q_{m-n}B_n(K)-B_0(\mathcal H)\sum_{m=0}^{\infty}Q_m\mathcal{M}_m(K,\mathcal{H})\Bigg]+O(\Omega^{2})\text{,}\label{eq:del1}
\end{align}
with
\begin{subequations}\label{eq:LMmdef}
\begin{align}
  \mathcal{L}_m(K,\mathcal{H})&=\sqrt{2K}\left(\frac{P_K}{P_0}\right)\int_0^{\mathcal{H}-K}\frac{\dd\eta}{\sqrt{\eta}}\left(\frac{\eta}{K}\right)^m\exp\left(-\frac{\eta}{H}\right)\text{,}\label{eq:Lmdef}\\
  \mathcal{M}_m(K,\mathcal{H})&=\sqrt{2K}\left(\frac{P_{\mathcal H}}{P_0}\right)\int_0^{\mathcal{H}-K}\frac{\dd\eta}{\sqrt{\eta}}\left(\frac{\eta}{K}\right)^m\text{,}\label{eq:Mmdef}
\end{align}
\end{subequations}
and
\begin{equation}
  m_d=\left\{
  \begin{array}{l l}
    m & \mathrm{for}\ m\leqslant d\text{,}\\
    d & \mathrm{for}\ m>d\text{.}
  \end{array}
  \right.
\end{equation}

Each $\mathcal L_m$ and $\mathcal M_m$ can now be integrated exactly. The solutions for the first terms (i.e. $m=0$) are given by
\begin{subequations}\label{eq:L0M0}
\begin{align}
  \mathcal{L}_0(K,\mathcal H)&=\sqrt{2\pi}\sqrt{HK}\exp\left(-\frac{K-R}{H}\right)\mathrm{erf}\left(\sqrt{\frac{\mathcal{H}-K}{H}}\right)\text{,}\label{eq:L0}\\
  \mathcal{M}_0(K,\mathcal{H})&=2\sqrt{2K}\sqrt{\mathcal{H}-K}\exp\left(-\frac{\mathcal{H}-R}{H}\right)\text{,}\label{eq:M0}
\end{align}
\end{subequations}
where $\mathrm{erf}(x)$ denotes the well-known error function
\begin{equation}
  \mathrm{erf}(x)=\frac{2}{\sqrt{\pi}}\int_0^x\exp\left(-y^2\right)\dd y\text{.}
\end{equation}

The following solutions (i.e. $m\geqslant 1$) are conveniently expressed in terms of $\mathcal L_0$, $\mathcal M_0$, and the $m$th power of $H/K$:
\begin{subequations}\label{eq:LMmLM}
\begin{align}
  \mathcal{L}_m(K,\mathcal{H})&=\frac{(2m-1)!!}{2^m}\left(\frac{H}{K}\right)^m\left[\mathcal{L}_0-\mathcal{M}_0\sum_{p=0}^{m-1}\frac{2^p}{(2p+1)!!}\left(\frac{\mathcal{H}-K}{H}\right)^p\right]\text{,}\label{eq:LmLM}\\
  \mathcal{M}_m(K,\mathcal{H})&=\frac{\mathcal{M}_0}{(2m+1)}\left(\frac{H}{K}\right)^m\left(\frac{\mathcal H-K}{H}\right)^m\text{.}\label{eq:MmLM}
\end{align}
\end{subequations}

It is seen from \eqref{eq:L0M0} that both $\mathcal{L}_0$ and $\mathcal{M}_0$ vanish when $K\to\mathcal H$. In addition, both $\mathcal{L}_0$ and $\mathcal{M}_0$ are only defined for $K\leqslant\mathcal H$. Indeed, $K>\mathcal H$ would return a non-physical imaginary number. This fact states that the refractive delay due to the optical medium is only observed for an optical ray crossing through the refractive domain $\mathcal D$ as expected.

Finally, the expression for the first order delay function is inferred after substituting for $\mathcal L_m$ and $\mathcal M_m$ from \eqref{eq:LMmLM} into \eqref{eq:del1} which eventually returns
\begin{align}
  \Delta^{(1)}(K,\mathcal H)&=C^2\left(\Delta^{(1)}_{\mathcal L_0}+\Delta^{(1)}_{\mathcal M_0}\right)_{(K,\mathcal H)}+O(\Omega^2)\text{,}
  \label{eq:del1HK}
\end{align}
where
\begin{subequations}\label{eq:del1abc}
\begin{align}
  \Delta^{(1)}_{\mathcal L_0}(K,\mathcal H)&=\mathcal{L}_0\sum_{m=0}^{\infty}\frac{(2m-1)!!}{2^m}\left(\frac{H}{K}\right)^m\sum_{n=0}^{m_d}Q_{m-n}B_n(K)\text{,}\label{eq:del1a}
\end{align}
\begin{align}
  \Delta^{(1)}_{\mathcal M_0}(K,\mathcal H)&=-\mathcal{M}_0\Bigg[\sum_{m=1}^{\infty}\frac{(2m-1)!!}{2^m}\left(\frac{H}{K}\right)^m\sum_{p=0}^{m-1}\frac{2^p}{(2p+1)!!}\left(\frac{\mathcal{H}-K}{H}\right)^p\sum_{n=0}^{m_d}Q_{m-n}B_n(K)\nonumber\\
  &+B_0(\mathcal H)\sum_{m=0}^{\infty}\frac{Q_m}{(2m+1)}\left(\frac{H}{K}\right)^m\left(\frac{\mathcal H-K}{H}\right)^m\Bigg]\text{.}\label{eq:del1b}
\end{align}
\end{subequations}

After substituting for $K$, $C^2$, and $\bm S_{AB}$ from \eqref{eq:K}, \eqref{eq:C2}, and \eqref{eq:sig}, into \eqref{eq:L0M0}, \eqref{eq:del1HK}, and \eqref{eq:del1abc}, we find the expression for the delay function in term of $\bm x_A$ and $\bm x_B$ as it is usually done in the literature about time transfer functions. We eventually get a relationship as follows
\begin{equation}
  \mathcal{T}(\bm x_A,\bm x_B)=\frac{\Vert\bm x_B-\bm x_A\Vert}{c}+\frac{N_0}{c}\Delta^{(1)}(\bm x_A,\bm x_B)+O(N_0^2)\text{.}
\end{equation}

We recall that the global frame is centered at the occulting body center of mass and is non-rotating with respect to distant stars. The atmosphere is still in the frame attached to the occulting body, namely the rotating frame. The refractivity profile is given by \eqref{eq:Nh}, where the pressure profile is an exponential function and where the temperature profile is a polynomial function of arbitrary degree $d$.

\subsection{Frequency transfer}
\label{sec:FTRO}

Hereafter, we focus on the determination of the frequency transfer. After inserting \eqref{eq:RTFdecR} and \eqref{eq:delRrPMN0G} into \eqref{eq:qTF}, we deduce
\begin{equation}
  q_A=1+\bm \beta_A\cdot\underbar{$\bm l$}_A\text{,} \qquad q_B=1+\bm \beta_B\cdot\underbar{$\bm l$}_B\text{,}\label{eq:qAqB}
\end{equation}
where the components of the covectors $\underbar{$\bm l$}_A$ and $\underbar{$\bm l$}_B$ have been introduced such as
\begin{subequations}\label{eq:lAlB}
\begin{align}
  \underbar{$\bm l$}_A(\bm x_A,\bm x_B,N_0)&=-\bm N_{AB}+\sum_{m=1}^{\infty}(N_0)^m\underbar{$\bm l$}_A^{(m)}(\bm x_A,\bm x_B)\text{,}\\
  \underbar{$\bm l$}_B(\bm x_A,\bm x_B,N_0)&=-\bm N_{AB}+\sum_{m=1}^{\infty}(N_0)^m\underbar{$\bm l$}_B^{(m)}(\bm x_A,\bm x_B)\text{,}
\end{align}
\end{subequations}
with
\begin{subequations}\label{eq:lA1lB1}
\begin{align}
  \underbar{$\bm l$}_A^{(m)}(\bm x_A,\bm x_B)&=\left[\frac{\partial\Delta^{(m)}}{\partial\bm x_A}\right]_{(\bm x_A,\bm x_B)}\text{,}\\
  \underbar{$\bm l$}_B^{(m)}(\bm x_A,\bm x_B)&=-\left[\frac{\partial\Delta^{(m)}}{\partial\bm x_B}\right]_{(\bm x_A,\bm x_B)}\text{.}
\end{align}
\end{subequations}

In order to derive the explicit expression for the frequency transfer, we need to determine $\underbar{$\bm l$}_A$ and $\underbar{$\bm l$}_B$. To do so we have to find the derivative of the time delay with respect to the impact parameter, and then the derivative of $K$ with respect to the components of the position vectors at the level of the emission and reception. The latters are easily inferred from \eqref{eq:K}. After inserting the so-obtained relations into \eqref{eq:lA1lB1}, we end up with
\begin{subequations}\label{eq:lAlB1}
\begin{align}
  \underbar{$\bm l$}_A^{(m)}(\bm x_A,\bm x_B)&=\bm n_K\left(\frac{\bm N_{AB}\cdot\bm x_B}{R_{AB}}\right)\left[\frac{\partial\Delta^{(m)}}{\partial K}\right]_{(\bm x_A,\bm x_B)}\text{,}\label{eq:lA}\\
  \underbar{$\bm l$}_B^{(m)}(\bm x_A,\bm x_B)&=-\bm n_K\left(1-\frac{\bm N_{AB}\cdot\bm x_B}{R_{AB}}\right)\left[\frac{\partial\Delta^{(m)}}{\partial K}\right]_{(\bm x_A,\bm x_B)}\text{.}\label{eq:lB}
\end{align}
\end{subequations}

Eqs. \eqref{eq:lAlB1} together with \eqref{eq:nK} and \eqref{eq:nKNAB} show that the directions $\underbar{$\bm l$}_A$ and $\underbar{$\bm l$}_B$ are unit triples at the first post-Minkowskian order (i.e. $m=1$), namely
\begin{equation}
  \Vert\underbar{$\bm l$}_A\Vert=1+O(N_0^2)\text{,} \qquad \Vert\underbar{$\bm l$}_B\Vert=1+O(N_0^2)\text{.}
\end{equation}
In addition, we notice from \eqref{eq:lB} that the covector at reception can be written such as
\begin{equation}
  \underbar{$\bm l$}_B=\underbar{$\bm l$}_A-\bm n_K\sum_{m=1}^{\infty}(N_0)^m\left[\frac{\partial\Delta^{(m)}}{\partial K}\right]_{(\bm x_A,\bm x_B)}\text{,}
\end{equation}
which shows, after invoking \eqref{eq:nKxNAB}, that
\begin{equation}
  \underbar{$\bm l$}_A\times\underbar{$\bm l$}_B=-\bm S_{AB}\sum_{m=1}^{\infty}(N_0)^m\left[\frac{\partial\Delta^{(m)}}{\partial K}\right]_{(\bm x_A,\bm x_B)}\text{,}
\end{equation}
These relations are useful for defining the bending angle $\phi$. Usually, it is introduced using the scalar product between the two tangent vectors, however in order to avoid numerical errors especially when dealing with small angles, it is more appropriate to introduce $\phi$ such as
\begin{equation}
  \phi(\bm x_A,\bm x_B)=\arcsin\left[\frac{\underbar{$\bm l$}_A\times\underbar{$\bm l$}_B}{\Vert\underbar{$\bm l$}_A\Vert\,\Vert\underbar{$\bm l$}_B\Vert}\cdot\bm S_{AB}\right]\text{.}
\end{equation}
Therefore, it is seen from Eqs. \eqref{eq:lAlB} that the bending angle assumes a post-Minkowskian expansion too, namely
\begin{equation}
  \phi(\bm x_A,\bm x_B,N_0)=\sum_{m=1}^\infty(N_0)^m\phi^{(m)}(\bm x_A,\bm x_B)\text{,}
  \label{eq:epsPM}
\end{equation}
where the first order term satisfies
\begin{equation}
  \phi^{(1)}(\bm x_A,\bm x_B)=-\Bigg[\frac{\partial\Delta^{(1)}}{\partial K}\Bigg]_{(\bm x_A,\bm x_B)}\text{.}
  \label{eq:epsN0}
\end{equation}

The last missing piece is now the derivative of the time delay with respect to the impact parameter. It can be derived from \eqref{eq:del1HK} and \eqref{eq:del1abc} as
\begin{align}
  \Bigg[\frac{\partial \Delta^{(1)}}{\partial K}\Bigg]_{(K,\mathcal H)}&=C^2\left(\frac{\partial \Delta^{(1)}_{\mathcal L_0}}{\partial K}+\frac{\partial \Delta^{(1)}_{\mathcal M_0}}{\partial K}\right)_{(K,\mathcal H)}-\frac{2D}{K}\left(\Delta^{(1)}_{\mathcal L_0}+\Delta^{(1)}_{\mathcal M_0}\right)_{(K,\mathcal H)}+O(\Omega^2)\text{,}\label{eq:Ddel}
\end{align}
where
\begin{subequations}\label{eq:DdelL0M0}
\begin{align}
  \Bigg[\frac{\partial \Delta^{(1)}_{\mathcal L_0}}{\partial K}\Bigg]_{(K,\mathcal H)}&=-\frac{\mathcal L_0}{H}\sum_{m=0}^{\infty}\frac{(2m-1)!!}{2^m}\left(\frac{H}{K}\right)^m\sum_{n=0}^{m_d}Q_{m-n}\sum_{l=n}^{d}\left(
  \begin{array}{c}
    l\\
    n
  \end{array}
  \right)b_lK^{l}\times\nonumber\\
  &\times\left[1+\left(\frac{H}{K}\right)\left(m-l-\frac{1}{2}+\frac{\mathcal{M}_0}{2\mathcal{L}_0}\frac{K}{\mathcal H-K}\right)\right]\text{,}
\end{align}
\begin{align}
  \Bigg[\frac{\partial \Delta^{(1)}_{\mathcal M_0}}{\partial K}\Bigg]_{(K,\mathcal H)}&=\frac{\mathcal M_0}{K}\Bigg\{\sum_{m=1}^{\infty}\frac{(2m-1)!!}{2^m}\left(\frac{H}{K}\right)^m\sum_{n=0}^{m_d}Q_{m-n}\sum_{l=n}^{d}\left(
  \begin{array}{c}
    l\\
    n
  \end{array}
  \right)b_lK^{l}\times\nonumber\\
  &\times\sum_{p=0}^{m-1}\frac{2^p}{(2p+1)!!}\left(\frac{\mathcal H-K}{H}\right)^p\left[\frac{2(p-m+l+1)K+(2m-2l-1)\mathcal H}{2(\mathcal H-K)}\right]\nonumber\\
  &+B_0(\mathcal H)\sum_{m=0}^{\infty}\frac{Q_m}{(2m+1)}\left(\frac{H}{K}\right)^m\left(\frac{\mathcal H-K}{H}\right)^m\left[\frac{2K+(2m-1)\mathcal H}{2(\mathcal H-K)}\right]\Bigg\}\text{.}
\end{align}
\end{subequations}

After substituting for $K$, $C^2$, $D$, and $\bm S_{AB}$ from \eqref{eq:K}, \eqref{eq:C2}, \eqref{eq:D}, and \eqref{eq:sig} into \eqref{eq:Ddel} and \eqref{eq:DdelL0M0}, we find expressions for the derivatives in terms of $\bm x_A$ and $\bm x_B$ as it is usually done in the literature about time transfer functions. The expressions for $\underbar{$\bm l$}_A$ and $\underbar{$\bm l$}_B$ are obtained after inserting \eqref{eq:Ddel} and \eqref{eq:DdelL0M0} into \eqref{eq:lAlB1} and \eqref{eq:lAlB}, and then the frequency transfer is simply given by \eqref{eq:qAqB} and \eqref{eq:dopTF}.

\subsection{Limits when $\mathcal H\to\infty$}

As mentioned earlier in Sec. \ref{sec:Nh}, in the context of radio occultation experiments, the effect of refractivity is often negligible when $K$ approaches $\mathcal H$. This is mainly due to the fast decrease of the exponential pressure profile when the value of the impact parameter increases.

We saw earlier that the refractive profile in Eq. \eqref{eq:Neta} has the expected limit when $\mathcal H\to\infty$. Hence, if one is interested in applications for occultation experiments one can safely replace $\mathcal{L}_0$ and $\mathcal{M}_0$ (in Eqs. \eqref{eq:L0M0}) by their following limits
\begin{subequations}
\begin{align}
  &\lim_{\mathcal H\to\infty}\mathcal{L}_0(K,\mathcal H)=\sqrt{2\pi}\sqrt{HK}\exp\left(-\frac{K-R}{H}\right)\text{,}\\
  &\lim_{\mathcal H\to\infty}\mathcal{M}_0(K,\mathcal H)=0\text{,}
\end{align}
\end{subequations}
meaning that all terms proportional to $\mathcal M_0$ vanish.

Therefore, the time transfer function simplifies to
\begin{align}
  \mathcal{T}(\bm x_A,\bm x_B)&=\frac{\Vert\bm x_B-\bm x_A\Vert}{c}\nonumber\\
  &+\frac{N_0}{c}\sqrt{2\pi}\sqrt{H}\sqrt{\Vert\bm N_{AB}\times\bm x_B\Vert}\exp\left(-\frac{\Vert\bm N_{AB}\times\bm x_B\Vert-R}{H}\right)\left[1+2\bm\Omega\cdot\frac{\bm N_{AB}\times\bm x_B}{\Vert\bm N_{AB}\times\bm x_B\Vert}+O(\Omega^{2})\right]\times\nonumber\\
  &\times\sum_{m=0}^{\infty}\frac{(2m-1)!!}{2^m}\left(\frac{H}{\Vert\bm N_{AB}\times\bm x_B\Vert}\right)^{m}\sum_{n=0}^{m_d}Q_{m-n}\sum_{l=n}^{d}\left(
  \begin{array}{c}
    l\\
    n
  \end{array}\right)b_l\Vert\bm N_{AB}\times\bm x_B\Vert^l+O(N_0^2)\text{.}\label{eq:TTFsimp}
\end{align}
Similarly, the covectors $\underbar{$\bm l$}_A$ and $\underbar{$\bm l$}_B$ are now given by
\begin{subequations}\label{eq:lAlBsimp}
\begin{align}
  \underbar{$\bm l$}_A(\bm x_A,\bm x_B)&=-\bm N_{AB}\nonumber\\
  &-N_0\sqrt{2\pi}\sqrt{\frac{\Vert\bm N_{AB}\times\bm x_B\Vert}{H}}\exp\left(-\frac{\Vert\bm N_{AB}\times\bm x_B\Vert-R}{H}\right)\times\nonumber\\
  &\times\sum_{m=0}^{\infty}\frac{(2m-1)!!}{2^m}\left(\frac{H}{\Vert\bm N_{AB}\times\bm x_B\Vert}\right)^m\sum_{n=0}^{m_d}Q_{m-n}\sum_{l=n}^{d}\left(
  \begin{array}{c}
    l\\
    n
  \end{array}
  \right)b_l\Vert\bm N_{AB}\times\bm x_B\Vert^{l}\times\nonumber\\
  &\times\Bigg\{1+\frac{H}{\Vert\bm N_{AB}\times\bm x_B\Vert}\left(m-l-\frac{1}{2}\right)+2\bm\Omega\cdot\frac{\bm N_{AB}\times\bm x_B}{\Vert\bm N_{AB}\times\bm x_B\Vert}\left[1+\frac{H}{\Vert\bm N_{AB}\times\bm x_B\Vert}\left(m-l-\frac{3}{2}\right)\right]+O(\Omega^2)\Bigg\}\times\nonumber\\
  &\times\Bigg.\left(\frac{\bm N_{AB}\cdot\bm x_B}{R_{AB}}\right)\frac{\bm N_{AB}\times\bm x_B}{\Vert\bm N_{AB}\times\bm x_B\Vert}\times\bm N_{AB}+O(N_0^2)\text{,}
\end{align}
and
\begin{align}
  \underbar{$\bm l$}_B(\bm x_A,\bm x_B)&=-\bm N_{AB}\nonumber\\
  &+N_0\sqrt{2\pi}\sqrt{\frac{\Vert\bm N_{AB}\times\bm x_B\Vert}{H}}\exp\left(-\frac{\Vert\bm N_{AB}\times\bm x_B\Vert-R}{H}\right)\times\nonumber\\
  &\times\sum_{m=0}^{\infty}\frac{(2m-1)!!}{2^m}\left(\frac{H}{\Vert\bm N_{AB}\times\bm x_B\Vert}\right)^m\sum_{n=0}^{m_d}Q_{m-n}\sum_{l=n}^{d}\left(
  \begin{array}{c}
    l\\
    n
  \end{array}
  \right)b_l\Vert\bm N_{AB}\times\bm x_B\Vert^{l}\times\nonumber\\
  &\times\Bigg\{1+\frac{H}{\Vert\bm N_{AB}\times\bm x_B\Vert}\left(m-l-\frac{1}{2}\right)+2\bm\Omega\cdot\frac{\bm N_{AB}\times\bm x_B}{\Vert\bm N_{AB}\times\bm x_B\Vert}\left[1+\frac{H}{\Vert\bm N_{AB}\times\bm x_B\Vert}\left(m-l-\frac{3}{2}\right)\right]+O(\Omega^2)\Bigg\}\times\nonumber\\
  &\times\Bigg.\left(1-\frac{\bm N_{AB}\cdot\bm x_B}{R_{AB}}\right)\frac{\bm N_{AB}\times\bm x_B}{\Vert\bm N_{AB}\times\bm x_B\Vert}\times\bm N_{AB}+O(N_0^2)\text{.}
\end{align}
\end{subequations}

The expression for the frequency transfer is directly inferred after inserting these last two expressions into \eqref{eq:qAqB} while making use of \eqref{eq:dopTF}.

\section{Numerical ray-tracing}
\label{sec:NumRT}

In this section, we perform a numerical integration of the equations for optical rays toward a spherically symmetric planetary atmosphere being rigidly rotating. We consider the case of an atmosphere with drastic changes in its temperature profile. We simulate the time and frequency transfers for a one-way downlink between an emitter in Keplerian orbit around the occulting planet and a receiver at infinity. We compare the numerical results to analytical solutions derived in Sec.~\ref{sec:mathmod}.

\subsection{Optical rays equations}

The equations for optical rays propagating in a nondispersive isotropic medium have been derived in Sec. \ref{sec:relopt} (see Eqs. \eqref{eq:RTdef}). However, they can be further simplified. Indeed, we have assumed that the optical spacetime is spherically symmetric, constant, and stationary. Accordingly, the time component of the 4-wave vector is a first integral since it remains constant during the propagation of the radio signal through $\mathcal D$. In addition, we have assumed that the velocity of the medium is small with respect to the speed of light in a vacuum so we may only consider terms up to the first order in $\omega/c$. With these simplifications, the equations for optical rays eventually read as follows
\begin{subequations}\label{eq:RT}
\begin{equation}
  \frac{\dd x^0}{\dd\ell}=n\text{,}
\end{equation}
and
\begin{align}
  \frac{\dd\bm x}{\dd\ell}&=\frac{1}{n}\left[-\underbar{$\bm l$}+\frac{\omega}{c}(n^2-1)\bm e\times\bm x\right]\text{,}\\
  \frac{\dd\underbar{$\bm l$}}{\dd\ell}&=-\bm\nabla n+\frac{\omega}{c}\frac{(n^2-1)}{n}\bm e\times\underbar{$\bm l$}\text{.}
\end{align}
\end{subequations}
We emphasize that these equations reduce to the classical set of equations of geometrical optics when $\omega/c\to 0$ (see Sec. 3.2.1 of \citet{1999prop.book.....B} and also Sec. 85 of \citet{1960ecm..book.....L}).

Let us assume that the global frame $(\bm e_X,\bm e_Y,\bm e_Z)$ is centered at the planet's center of mass and is non rotating with respect to distant stars. The $(\bm e_X,\bm e_Y)$-plane is chosen to coincides with the occulting planet's equator. The $\bm e_Z$-axis is aligned with the planet instantaneous axis of rotation, that is to say $\bm e_Z=\bm e$. For convenience, we consider the case of an emitter at infinity whose direction vector is lying in the equatorial plane, so that we can choose to define $\bm e_Y=-\bm N_{AB}$. The emitter's orbit is characterized by the usual set of Keplerian elements, namely $(a_A,e_A,\iota_A,\Omega_A,\omega_A,\tau_A)$. Values of the selected Keplerian elements are given in Tab.~\ref{tab:KepA}. Then, the Cartesian position and velocity of the emitter in the global frame are given by Eqs.~(3.40) and (3.41) of \cite{2014gravbookP}. By substituting for the Keplerian elements from Tab.~\ref{tab:KepA} into Eq. (3.44) of \cite{2014gravbookP}, it is seen that the direction of the pericenter coincides with $\bm e_Y$-axis.

\begin{table}[b]
  \centering
  \caption[]{Values of Keplerian elements of the emitter.}
  \label{tab:KepA}
  \begin{tabular}{l c r}
    \hline\hline
    \multicolumn{1}{c}{Element} & Unit & \multicolumn{1}{c}{Value} \\
    \hline
    $a_A$ & $R$ & 2 \\
    $e_A$ & - & 0.1 \\
    $\iota_A$ & deg & -45 \\
    $\Omega_A$ & deg & 90 \\
    $\omega_A$ & deg & 0 \\
    $\tau_A$ & min & 50 \\
    \hline
  \end{tabular}
\end{table}

\begin{figure}
  \centering
  \vspace{0.2cm}
  \includegraphics[trim={0 0 0 0},clip]{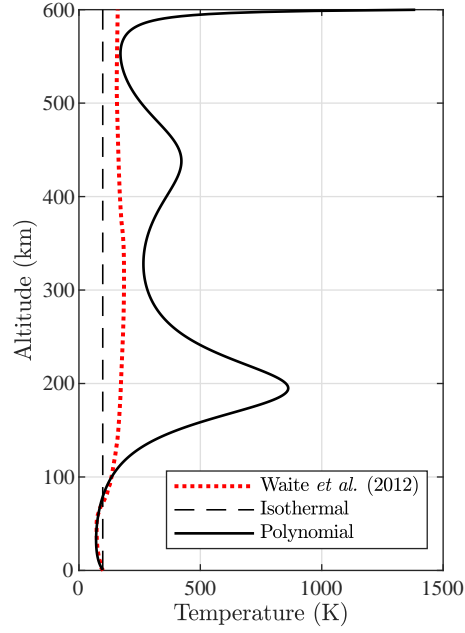}
  \caption{Temperature profiles inside the occulting planet's atmosphere. The plain curve represents a $6$th degree polynomial temperature profile, the dashed line represents an isothermal atmospheric profile, and the dotted line is Titan's atmospheric model of \cite{2012LPI....43.1232W}.}
  \label{fig:temp}
\end{figure}

We consider that the mass and radii of the occulting planet are similar to Titan's, that is to say $M=1.35\times10^{23}\ \mathrm{kg}$ and $R=2\,574\ \mathrm{km}$. However, in order to truly assess the accuracy of analytical solutions, we consider more extreme atmospheric physical properties than Titan's atmospheric model of \cite{2012LPI....43.1232W}. For instance, we would rather consider a completely made up temperature profile exhibiting high vertical gradients as shown in Fig.~\ref{fig:temp} (see plain curve). In addition, we assume that the atmosphere is fastly rotating with an angular rate of $2\pi\ \mathrm{rad}\cdot\mathrm{s}^{-1}$, so that the relativistic light-dragging term at the surface reaches $\Omega=0.05$.

The equations for optical rays are numerically integrated considering the following spherically symmetric index of refraction profile inside the domain $\mathcal D$ (see discussion in Sec. \ref{sec:Nh})
\begin{subequations}
\begin{equation}
  n(r)=1+N_0\exp\left(-\frac{r-R}{H}\right)\sum_{m=0}^db_mr^m-N_{\mathcal H}\text{,}
  \label{eq:nrSim}
\end{equation}
with $r=\Vert\bm x\Vert$ and
\begin{equation}
  N_{\mathcal H}=N_0\exp\left(-\frac{\mathcal H-R}{H}\right)\sum_{m=0}^db_m\mathcal H^m\text{.}
\end{equation}
\end{subequations}
The values of polynomial coefficients $b_m$ are given in Tab. \ref{tab:bm} for $d=6$. The size of the domain $\mathcal D$ is taken to be $\mathcal H=3\,174\ \mathrm{km}$, corresponding to an atmosphere thickness of $600\ \mathrm{km}$. We consider a scale height of $H=20\ \mathrm{km}$ (similar to Titan's), and use values of $N_0$ (post-Minkowskian parameter of the theory) ranging from $10^{-3}$ to $10^{-6}$ for illustration purposes. 

\begin{table}[b]
  \centering
  \caption[]{Values of $b_m$ coefficients for the determination of a degree 6 polynomial temperature profile.}
  \label{tab:bm}
  \begin{tabular}{c c l}
    \hline\hline
    Coefficient & Unit & \multicolumn{1}{c}{Value} \\
    \hline
    $b_{0}$ & - & $-5.415\,049\,754\,779\times 10^6$ \\
    $b_{1}$ & $\mathrm{km}^{-1}$ & $+1.132\,607\,910\,442\times 10^4$ \\
    $b_{2}$ & $\mathrm{km}^{-2}$ & $-9.860\,328\,832\,788\times 10^0$ \\
    $b_{3}$ & $\mathrm{km}^{-3}$ & $+4.573\,547\,412\,562\times 10^{-3}$ \\
    $b_{4}$ & $\mathrm{km}^{-4}$ & $-1.192\,048\,581\,350\times 10^{-6}$ \\
    $b_{5}$ & $\mathrm{km}^{-5}$ & $+1.655\,369\,690\,809\times 10^{-10}$ \\
    $b_{6}$ & $\mathrm{km}^{-6}$ & $-9.568\,664\,414\,388\times 10^{-15}$ \\
    \hline
  \end{tabular}
\end{table}

The equations for optical rays are only integrated inside the refractive domain $\mathcal D$. Outside, optical rays are simply assumed to propagate along straight lines according to the assumption \eqref{eq:flat}.

\subsection{Numerical integration and initial pointing}

The initial pointing direction at the level of the emitter is first assumed to be
\begin{equation}
  \underbar{$\bm l$}_A=-\bm N_{AB}\text{.}
  \label{eq:pointinit}
\end{equation}
The spacetime coordinates of the ray entrance point-event inside the atmosphere can always be determined once the initial pointing direction is known. Let $(ct_E,\bm x_E)$ be the coordinates of the entrance point-event $x_E$. It is clear that both $t_E$ and $\bm x_E$ are functions of $\underbar{$\bm l$}_A$, that is to say $t_E=t_E(\underbar{$\bm l$}_A)$ and $\bm x_E=\bm x_E(\underbar{$\bm l$}_A)$.

The initial pointing direction in \eqref{eq:pointinit} implies that $x_E\equiv x_+$ (see Eqs.~\eqref{eq:pe+-}) for the first iteration. Then, having the coordinates of $x_E$, the equations for optical rays in \eqref{eq:RT} can now be numerically integrated starting with $\ell_E=c(t_E-t_A)$ (we define $t_A$ as the origin of the coordinate time for the numerical integration) and
\begin{subequations}\label{eq:NumE}
\begin{align}
  x^0(\ell_E,\underbar{$\bm l$}_A)&=c(t_E-t_A)\text{,}\\
  \bm x(\ell_E,\underbar{$\bm l$}_A)&=\bm x_E\text{,}\\
  \underbar{$\bm l$}(\ell_E,\underbar{$\bm l$}_A)&=\underbar{$\bm l$}_A\text{.}
\end{align}
\end{subequations}
The numerical integration is stopped when the optical ray crosses the refractive domain $\mathcal D$ on the opposite side with respect to the entrance point-event $x_E$. Let $(ct_F,\bm x_F)$ be the coordinates of $x_F$, the final point-event for the numerical integration, that is to say
\begin{subequations}\label{eq:NumF}
\begin{align}
  x^0(\ell_F;\underbar{$\bm l$}_A)&=c(t_F-t_A)\text{,}\\
  \bm x(\ell_F;\underbar{$\bm l$}_A)&=\bm x_F\text{,}\\
  \underbar{$\bm l$}(\ell_F;\underbar{$\bm l$}_A)&=\underbar{$\bm l$}_F\text{.}
\end{align}
\end{subequations}
It is clear that coordinates of $x_F$ depend on the initial conditions that have been used for conducting the numerical integration, namely $\underbar{$\bm l$}_A$.

In general, because refractivity in $\mathcal D$ causes the optical ray to depart from its original direction, the direction of the ray at the exit of the atmosphere does not match the direction of the receiver. In order to make the two directions coincide, the initial pointing is iteratively corrected using a Newton-Raphson method \citep{1992nrfa.book.....P} for finding $\underbar{$\bm l$}_A$ from the following condition
\begin{equation}
  \underbar{$\bm l$}(\ell_F;\underbar{$\bm l$}_A)+\bm N_{AB}=\bm 0\text{.}
  \label{eq:NewRaph}
\end{equation}

Eqs. \eqref{eq:RT} are numerically integrated between \eqref{eq:NumE} and \eqref{eq:NumF} assuming a relative numerical error tolerance of $10^{-12}$. The components of $\underbar{$\bm l$}_A$ are iteratively determined from \eqref{eq:NewRaph} with the exact same accuracy. The partial derivatives of $\underbar{$\bm l$}$ with respect to $\underbar{$\bm l$}_A$, that are needed for solving the initial pointing from Eq. \eqref{eq:NewRaph}, are determined using a second order finite difference method. Finally, the time and frequency transfers are computed from numerical solutions of $t_F$, $\bm x_F$, and $\underbar{$\bm l$}_A$.

The well-known atmospheric delay is given by the difference between the total light-time needed for reaching $x_F$ from $x_A$ and the projection of $\bm x_F-\bm x_A$ along $\bm N_{AB}$, namely
\begin{equation}
  \Delta_{\mathrm{atm}}=t_F-t_A-\frac{(\bm x_F-\bm x_A)\cdot\bm N_{AB}}{c}\text{.}
  \label{eq:Delatm}
\end{equation}
The expression for the relative Doppler frequency shift due to the atmosphere, and for the case of an observer at infinity, is obtained after inserting $\underbar{$\bm l$}_B=-\bm N_{AB}$ into \eqref{eq:dopTF} and \eqref{eq:qAqB}, which eventually returns
\begin{equation}
  \frac{\Delta\nu_B}{[\nu_B]_{\mathrm{vac}}}=\frac{\nu_B-[\nu_B]_{\mathrm{vac}}}{[\nu_B]_{\mathrm{vac}}}=-\cfrac{\bm\beta_A\cdot(\underbar{$\bm l$}_A+\bm N_{AB})}{1+\bm\beta_A\cdot\underbar{$\bm l$}_A}\text{,}\label{eq:Dopatm}
\end{equation}
where $[\nu_B]_{\mathrm{vac}}$ is the frequency that would be observed at $x_B$ if the signal were to be transmitted in a neat vacuum. The expression for $[\nu_B]_{\mathrm{vac}}$ is inferred from Eqs. \eqref{eq:DopVac}:
\begin{equation}
  [\nu_B]_{\mathrm{vac}}=\nu_A\left[\frac{\nu_B}{\nu_A}\right]_{\mathrm{vac}}\text{.}
\end{equation}

Expression \eqref{eq:Dopatm} is used for computing the analytical relative Doppler frequency shift as well, where $\underbar{$\bm l$}_A$ is given by the analytical expression instead of the numerical one.

\subsection{Accuracy of analytical solutions}

We are now able to perform the comparison between analytical and numerical solutions for the time and frequency transfers considering the index of refraction profile given in \eqref{eq:nrSim}.

The analytical solution for the atmospheric time delay is constructed from the first order delay function in \eqref{eq:del1HK}, that is $N_0\Delta^{(1)}(K,\mathcal H)$. The summations over the index $m$ are stopped for $m=10$, but could have been stopped way before thanks to the small value of the coefficient for the expansion (e.g. for $K=R$, we have $H/K\simeq 0.008$). The analytical solution for the relative Doppler frequency shift is built from \eqref{eq:Dopatm} with $\underbar{$\bm l$}_A$ deduced from the first order terms in Eqs.~\eqref{eq:lAlB}, \eqref{eq:lAlB1}, and for $m=10$ in Eq.~\eqref{eq:Ddel}. Because the receiver is at infinity, we replace $\bm x_B$ by
\begin{equation}
  \bm x_B=\bm x_A+R_{AB}\bm N_{AB}
\end{equation}
any time it appears into analytical expressions.

\begin{figure*}
  \centering
  \vspace{0.2cm}
  \includegraphics[trim={0 0 0 0},clip]{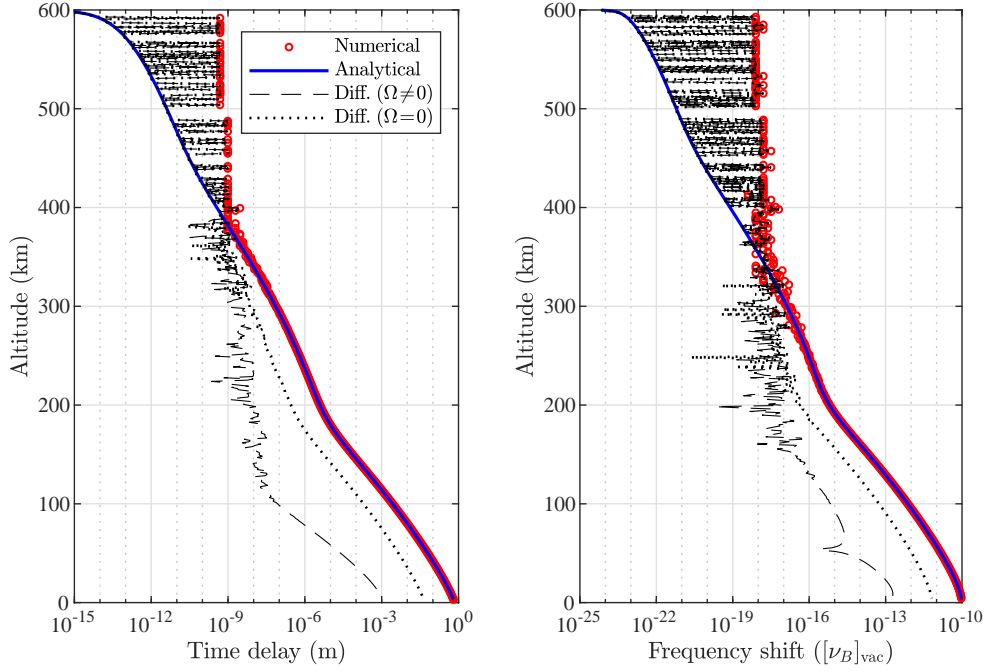}
  \caption{Time delay (left panel) and frequency shift (right panel) due to the atmosphere of the occulting planet for an index of refraction profile given in \eqref{eq:nrSim} with $N_0=10^{-6}$. Red circles represent results of the numerical integration (i.e. $\Delta_{\mathrm{atm}}$ in \eqref{eq:Delatm} for the left panel, and Eq. \eqref{eq:Dopatm} with $\underbar{$\bm l$}_A$ deduced from \eqref{eq:NewRaph} for the right panel). The thick blue line corresponds to the analytical predictions (i.e. $N_0\Delta^{(1)}(K,\mathcal H)$ in \eqref{eq:del1HK} for the left panel, and Eq. \eqref{eq:Dopatm} with $\underbar{$\bm l$}_A$ deduced from \eqref{eq:lAlB}, \eqref{eq:lAlB1}, and \eqref{eq:Ddel} for the right panel). The dashed lines correspond to the difference between numerical and analytical results. The dotted lines represent the same difference setting $\Omega=0$ into analytical solutions.}
  \label{fig:N0m6}
\end{figure*}

\begin{figure*}
  \centering
  \vspace{0.2cm}
  \includegraphics[trim={0 0 0 0},clip]{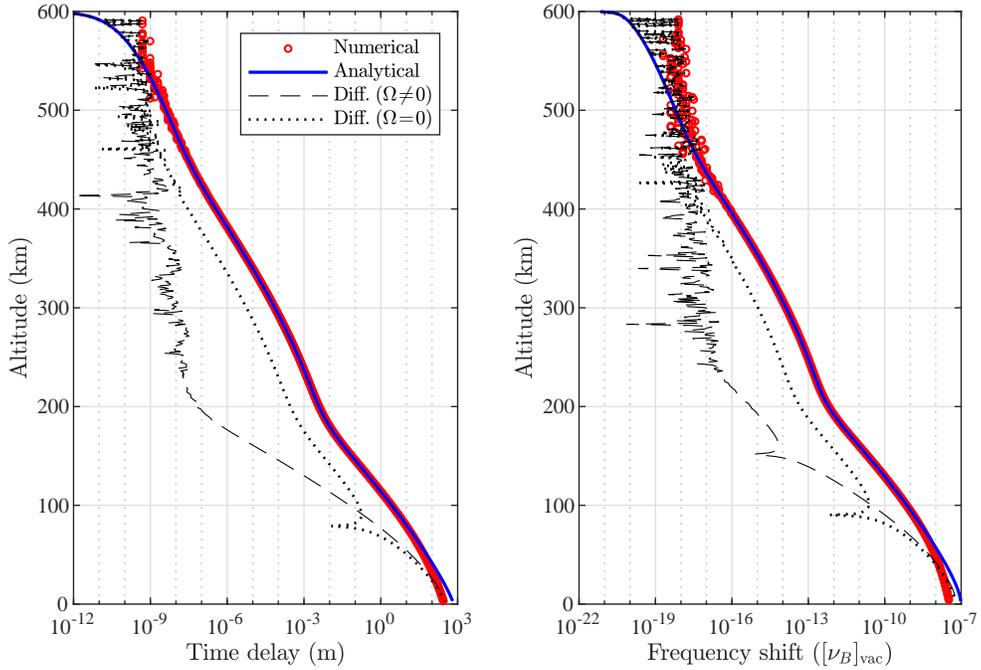}
  \caption{Time delay (left panel) and frequency shift (right panel) due to the atmosphere of the occulting planet for an index of refraction profile given in \eqref{eq:nrSim} with $N_0=10^{-3}$. The reader is referred to the caption of Fig. \ref{fig:N0m6} for more details.}
  \label{fig:N0m3}
\end{figure*}

Results of the comparison are presented in Fig. \ref{fig:N0m6} and \ref{fig:N0m3} for $N_0=10^{-6}$ and $10^{-3}$, respectively. In both figures, we notice that analytical solutions for the time and the frequency transfers succeed in reproducing perfectly the effects due to the vertical temperature variations (see e.g. signature at $h\simeq200\ \mathrm{km}$ in Figs. \ref{fig:temp}, \ref{fig:N0m6}, and \ref{fig:N0m3}). As a matter of fact, the difference between the numerical and the analytical profiles remains at the level of the numerical noise as shown in Fig. \ref{fig:N0m6}.

In addition, in order to assess the legitimacy of the light-dragging effect we show, on each plot, the difference of the numerical solution with an additional analytical solution built by setting $\Omega=0$. It is clearly seen that neglecting the light-dragging effect drastically decreases the accuracy of the analytical solution (up to 3 orders of magnitude for the time delay and up to 2 orders of magnitude for the relative Doppler frequency shift).

In this work, we focused our attention on the explicit resolution of the first post-Minkowskian order (see Eq. \eqref{eq:delPM1}) which corresponds to the well-known excess path delay when the velocity of the medium is neglected. The second order term is composed of a second order correction to the excess path delay and a geometric delay involving the derivative of the first order delay function. The effect of these neglected second order terms can be seen in Fig.~\ref{fig:N0m6} and \ref{fig:N0m3} when the differences between the numerical and analytical solutions start to increase. For $N_0=10^{-6}$ in Fig.~\ref{fig:N0m6}, the second order effects show up starting from $h\simeq 100\ \mathrm{km}$ and their influence increases when the altitude decreases, while they manifest at higher altitude, around $h\simeq 200\ \mathrm{km}$, for $N_0=10^{-3}$ in Fig.~\ref{fig:N0m3}.

For $N_0=10^{-3}$, the differences between analytical solutions and numerical results are thus dominated by numerical noise above $h\simeq 200\ \mathrm{km}$ and by the neglected second order below. The relative error has its minimum value of $0.001\%$ for $h\simeq 200\ \mathrm{km}$ and exceeds the $10\%$ level below $h\simeq 50\ \mathrm{km}$ for both the time delay and the relative Doppler frequency shift. This means that for high refractivity (i.e. $N_0=10^{-3}$) and for low altitude (i.e. $h<50\ \mathrm{km}$), we cannot expect the first order analytical solutions to describe the overall atmospheric effects with a relative accuracy better than one part in 10. In order to achieve a more accurate modeling, second order terms (i.e. the second order correction to the excess path delay and the geometric delay) shall be considered.

For $N_0=10^{-6}$ the relative error is dominated by numerical noise above $h\simeq 100\ \mathrm{km}$ and by the neglected second order below. For $h\simeq 100\ \mathrm{km}$ the relative error is $0.001\%$ and reaches $0.1\%$ at the ground level for both the time delay and the relative Doppler frequency shift. This means that for small refractivity (i.e. $N_0=10^{-6}$), neglecting second order terms would not make the relative errors larger than one part in $10^3$ on the atmospheric time delay and the relative Doppler frequency shift retrievals. In that respect, the maximum absolute errors (at the ground level) due to neglected second order terms is expected to be at the level of $1\ \mathrm{mm}$ on the time delay and $10^{-13}\ [\nu_B]_{\mathrm{vac}}$ on the Doppler frequency shift.

\section{Conclusions}
\label{sec:ccl}

In this work we presented a fully covariant analysis for deriving analytical expressions for the time/frequency transfers in the context of atmospheric occultation experiments. We combined two distinct relativistic theoretical tools, namely the Gordon's optical metric and the time transfer functions formalism. The first one is used to handle refractivity as spacetime curvature while the second one offers an efficient basis for sorting post-Minkowskian orders and for modeling time/frequency transfers in curved spacetime. We provided the integral form of the refractive delay function for any post-Minkowskian order and considered the case of an occultation by a steady rotating and spherically symmetric atmosphere. We assumed a refractivity profile driven by an exponential pressure profile and a polynomial temperature profile of arbitrary degree. We explicitly solved for the time/frequency transfers at first post-Minkowskian order in the limit where the angular velocity of the optical medium is small with respect to the speed of light in a vacuum. Finally, we assessed the accuracy of the first order analytical solutions by comparing them to results of a numerical integration of the equations for optical rays. We emphasized how complete these first order analytical solutions actually are. Indeed, they are able to properly consider any vertical temperature gradients and properly account for light-dragging effect due to the motion of the optical medium. We also noticed that for refractivity higher than $10^{-3}$, solutions that include up to the second post-Minkowskian order should be considered. The fully covariant method described in this paper can easily be extended in order to include following order even beyond spherical symmetry. An immediate application of the analytical method presented in this paper is the assessment of the expected sensitivities in pressure/density/temperature profiles for planetary atmospheric radio occultation experiments. This goes beyond the goals of this paper and will be the subject of future research. 

\begin{acknowledgements}
  A.B., M.Z., L.G.C., and P.T. are grateful to the Italian Space Agency (ASI) for financial support through Agreement No. 2018-25-HH.0 in the context of ESA’s JUICE mission, and Agreement No. 2020-13-HH.0 in the context of TRIDENT's mission Phase A study.
\end{acknowledgements}

\bibliographystyle{apsrev4-1}
\bibliography{Occultations}

\begin{thebibliography}{38}%
\makeatletter
\providecommand \@ifxundefined [1]{%
 \@ifx{#1\undefined}
}%
\providecommand \@ifnum [1]{%
 \ifnum #1\expandafter \@firstoftwo
 \else \expandafter \@secondoftwo
 \fi
}%
\providecommand \@ifx [1]{%
 \ifx #1\expandafter \@firstoftwo
 \else \expandafter \@secondoftwo
 \fi
}%
\providecommand \natexlab [1]{#1}%
\providecommand \enquote  [1]{``#1''}%
\providecommand \bibnamefont  [1]{#1}%
\providecommand \bibfnamefont [1]{#1}%
\providecommand \citenamefont [1]{#1}%
\providecommand \href@noop [0]{\@secondoftwo}%
\providecommand \href [0]{\begingroup \@sanitize@url \@href}%
\providecommand \@href[1]{\@@startlink{#1}\@@href}%
\providecommand \@@href[1]{\endgroup#1\@@endlink}%
\providecommand \@sanitize@url [0]{\catcode `\\12\catcode `\$12\catcode
  `\&12\catcode `\#12\catcode `\^12\catcode `\_12\catcode `\%12\relax}%
\providecommand \@@startlink[1]{}%
\providecommand \@@endlink[0]{}%
\providecommand \url  [0]{\begingroup\@sanitize@url \@url }%
\providecommand \@url [1]{\endgroup\@href {#1}{\urlprefix }}%
\providecommand \urlprefix  [0]{URL }%
\providecommand \Eprint [0]{\href }%
\providecommand \doibase [0]{http://dx.doi.org/}%
\providecommand \selectlanguage [0]{\@gobble}%
\providecommand \bibinfo  [0]{\@secondoftwo}%
\providecommand \bibfield  [0]{\@secondoftwo}%
\providecommand \translation [1]{[#1]}%
\providecommand \BibitemOpen [0]{}%
\providecommand \bibitemStop [0]{}%
\providecommand \bibitemNoStop [0]{.\EOS\space}%
\providecommand \EOS [0]{\spacefactor3000\relax}%
\providecommand \BibitemShut  [1]{\csname bibitem#1\endcsname}%
\let\auto@bib@innerbib\@empty
\bibitem [{\citenamefont {Linet}\ and\ \citenamefont
  {Teyssandier}(2002)}]{2002PhRvD..66b4045L}%
  \BibitemOpen
  \bibfield  {author} {\bibinfo {author} {\bibfnamefont {B.}~\bibnamefont
  {Linet}}\ and\ \bibinfo {author} {\bibfnamefont {P.}~\bibnamefont
  {Teyssandier}},\ }\href {\doibase 10.1103/PhysRevD.66.024045} {\bibfield
  {journal} {\bibinfo  {journal} {\prd}\ }\textbf {\bibinfo {volume} {66}},\
  \bibinfo {pages} {024045} (\bibinfo {year} {2002})}\BibitemShut {NoStop}%
\bibitem [{\citenamefont {Synge}(1960)}]{SyngeBookGR}%
  \BibitemOpen
  \bibfield  {author} {\bibinfo {author} {\bibfnamefont {J.~L.}\ \bibnamefont
  {Synge}},\ }\href {https://ui.adsabs.harvard.edu/abs/1960rgt..book.....S}
  {\emph {\bibinfo {title} {Relativity: The General Theory}}}\ (\bibinfo
  {publisher} {North-Holland Publ. Co.},\ \bibinfo {address} {Amsterdam},\
  \bibinfo {year} {1960})\BibitemShut {NoStop}%
\bibitem [{\citenamefont {Le~Poncin-Lafitte}\ \emph {et~al.}(2004)\citenamefont
  {Le~Poncin-Lafitte}, \citenamefont {Linet},\ and\ \citenamefont
  {Teyssandier}}]{2004CQGra..21.4463L}%
  \BibitemOpen
  \bibfield  {author} {\bibinfo {author} {\bibfnamefont {C.}~\bibnamefont
  {Le~Poncin-Lafitte}}, \bibinfo {author} {\bibfnamefont {B.}~\bibnamefont
  {Linet}}, \ and\ \bibinfo {author} {\bibfnamefont {P.}~\bibnamefont
  {Teyssandier}},\ }\href {\doibase 10.1088/0264-9381/21/18/012} {\bibfield
  {journal} {\bibinfo  {journal} {Classical and Quantum Gravity}\ }\textbf
  {\bibinfo {volume} {21}},\ \bibinfo {pages} {4463} (\bibinfo {year}
  {2004})}\BibitemShut {NoStop}%
\bibitem [{\citenamefont {Teyssandier}\ and\ \citenamefont
  {Le~Poncin-Lafitte}(2008)}]{2008CQGra..25n5020T}%
  \BibitemOpen
  \bibfield  {author} {\bibinfo {author} {\bibfnamefont {P.}~\bibnamefont
  {Teyssandier}}\ and\ \bibinfo {author} {\bibfnamefont {C.}~\bibnamefont
  {Le~Poncin-Lafitte}},\ }\href {\doibase 10.1088/0264-9381/25/14/145020}
  {\bibfield  {journal} {\bibinfo  {journal} {Classical and Quantum Gravity}\
  }\textbf {\bibinfo {volume} {25}},\ \bibinfo {pages} {145020} (\bibinfo
  {year} {2008})}\BibitemShut {NoStop}%
\bibitem [{\citenamefont {Richter}\ and\ \citenamefont
  {Matzner}(1983)}]{1983PhRvD..28.3007R}%
  \BibitemOpen
  \bibfield  {author} {\bibinfo {author} {\bibfnamefont {G.~W.}\ \bibnamefont
  {Richter}}\ and\ \bibinfo {author} {\bibfnamefont {R.~A.}\ \bibnamefont
  {Matzner}},\ }\href {\doibase 10.1103/PhysRevD.28.3007} {\bibfield  {journal}
  {\bibinfo  {journal} {\prd}\ }\textbf {\bibinfo {volume} {28}},\ \bibinfo
  {pages} {3007} (\bibinfo {year} {1983})}\BibitemShut {NoStop}%
\bibitem [{\citenamefont {Brumberg}(1987)}]{1987KFNT....3....8B}%
  \BibitemOpen
  \bibfield  {author} {\bibinfo {author} {\bibfnamefont {V.~A.}\ \bibnamefont
  {Brumberg}},\ }\href {http://adsabs.harvard.edu/abs/1987KFNT....3....8B}
  {\bibfield  {journal} {\bibinfo  {journal} {Kinematika i Fizika Nebesnykh
  Tel}\ }\textbf {\bibinfo {volume} {3}},\ \bibinfo {pages} {8} (\bibinfo
  {year} {1987})}\BibitemShut {NoStop}%
\bibitem [{\citenamefont {Bourgoin}(2020)}]{PhysRevD.101.064035}%
  \BibitemOpen
  \bibfield  {author} {\bibinfo {author} {\bibfnamefont {A.}~\bibnamefont
  {Bourgoin}},\ }\href {\doibase 10.1103/PhysRevD.101.064035} {\bibfield
  {journal} {\bibinfo  {journal} {Phys. Rev. D}\ }\textbf {\bibinfo {volume}
  {101}},\ \bibinfo {pages} {064035} (\bibinfo {year} {2020})}\BibitemShut
  {NoStop}%
\bibitem [{\citenamefont {Gordon}(1923)}]{doi101002andp19233772202}%
  \BibitemOpen
  \bibfield  {author} {\bibinfo {author} {\bibfnamefont {W.}~\bibnamefont
  {Gordon}},\ }\href {\doibase 10.1002/andp.19233772202} {\bibfield  {journal}
  {\bibinfo  {journal} {Annalen der Physik}\ }\textbf {\bibinfo {volume}
  {377}},\ \bibinfo {pages} {421} (\bibinfo {year} {1923})}\BibitemShut
  {NoStop}%
\bibitem [{\citenamefont {{Kliore}}\ \emph {et~al.}(1965)\citenamefont
  {{Kliore}}, \citenamefont {{Cain}}, \citenamefont {{Levy}}, \citenamefont
  {{Eshleman}}, \citenamefont {{Fjeldbo}},\ and\ \citenamefont
  {{Drake}}}]{1965Sci...149.1243K}%
  \BibitemOpen
  \bibfield  {author} {\bibinfo {author} {\bibfnamefont {A.}~\bibnamefont
  {{Kliore}}}, \bibinfo {author} {\bibfnamefont {D.~L.}\ \bibnamefont
  {{Cain}}}, \bibinfo {author} {\bibfnamefont {G.~S.}\ \bibnamefont {{Levy}}},
  \bibinfo {author} {\bibfnamefont {V.~R.}\ \bibnamefont {{Eshleman}}},
  \bibinfo {author} {\bibfnamefont {G.}~\bibnamefont {{Fjeldbo}}}, \ and\
  \bibinfo {author} {\bibfnamefont {F.~D.}\ \bibnamefont {{Drake}}},\ }\href
  {\doibase 10.1126/science.149.3689.1243} {\bibfield  {journal} {\bibinfo
  {journal} {Science}\ }\textbf {\bibinfo {volume} {149}},\ \bibinfo {pages}
  {1243} (\bibinfo {year} {1965})}\BibitemShut {NoStop}%
\bibitem [{\citenamefont {{Fjeldbo}}\ and\ \citenamefont
  {{Eshleman}}(1965)}]{1965JGR....70.3217F}%
  \BibitemOpen
  \bibfield  {author} {\bibinfo {author} {\bibfnamefont {G.}~\bibnamefont
  {{Fjeldbo}}}\ and\ \bibinfo {author} {\bibfnamefont {V.~R.}\ \bibnamefont
  {{Eshleman}}},\ }\href {\doibase 10.1029/JZ070i013p03217} {\bibfield
  {journal} {\bibinfo  {journal} {\jgr}\ }\textbf {\bibinfo {volume} {70}},\
  \bibinfo {pages} {3217} (\bibinfo {year} {1965})}\BibitemShut {NoStop}%
\bibitem [{\citenamefont {{Fjeldbo}}\ and\ \citenamefont
  {{Eshleman}}(1968)}]{1968P&SS...16.1035F}%
  \BibitemOpen
  \bibfield  {author} {\bibinfo {author} {\bibfnamefont {G.}~\bibnamefont
  {{Fjeldbo}}}\ and\ \bibinfo {author} {\bibfnamefont {V.~R.}\ \bibnamefont
  {{Eshleman}}},\ }\href {\doibase 10.1016/0032-0633(68)90020-2} {\bibfield
  {journal} {\bibinfo  {journal} {\planss}\ }\textbf {\bibinfo {volume} {16}},\
  \bibinfo {pages} {1035} (\bibinfo {year} {1968})}\BibitemShut {NoStop}%
\bibitem [{\citenamefont {{Lindal}}\ \emph {et~al.}(1985)\citenamefont
  {{Lindal}}, \citenamefont {{Sweetnam}},\ and\ \citenamefont
  {{Eshleman}}}]{1985AJ.....90.1136L}%
  \BibitemOpen
  \bibfield  {author} {\bibinfo {author} {\bibfnamefont {G.~F.}\ \bibnamefont
  {{Lindal}}}, \bibinfo {author} {\bibfnamefont {D.~N.}\ \bibnamefont
  {{Sweetnam}}}, \ and\ \bibinfo {author} {\bibfnamefont {V.~R.}\ \bibnamefont
  {{Eshleman}}},\ }\href {\doibase 10.1086/113820} {\bibfield  {journal}
  {\bibinfo  {journal} {\aj}\ }\textbf {\bibinfo {volume} {90}},\ \bibinfo
  {pages} {1136} (\bibinfo {year} {1985})}\BibitemShut {NoStop}%
\bibitem [{\citenamefont {{Lindal}}\ \emph {et~al.}(1987)\citenamefont
  {{Lindal}}, \citenamefont {{Lyons}}, \citenamefont {{Sweetnam}},
  \citenamefont {{Eshleman}},\ and\ \citenamefont
  {{Hinson}}}]{1987JGR....9214987L}%
  \BibitemOpen
  \bibfield  {author} {\bibinfo {author} {\bibfnamefont {G.~F.}\ \bibnamefont
  {{Lindal}}}, \bibinfo {author} {\bibfnamefont {J.~R.}\ \bibnamefont
  {{Lyons}}}, \bibinfo {author} {\bibfnamefont {D.~N.}\ \bibnamefont
  {{Sweetnam}}}, \bibinfo {author} {\bibfnamefont {V.~R.}\ \bibnamefont
  {{Eshleman}}}, \ and\ \bibinfo {author} {\bibfnamefont {D.~P.}\ \bibnamefont
  {{Hinson}}},\ }\href {\doibase 10.1029/JA092iA13p14987} {\bibfield  {journal}
  {\bibinfo  {journal} {\jgr}\ }\textbf {\bibinfo {volume} {92}},\ \bibinfo
  {pages} {14987} (\bibinfo {year} {1987})}\BibitemShut {NoStop}%
\bibitem [{\citenamefont {{Lindal}}(1992)}]{1992AJ....103..967L}%
  \BibitemOpen
  \bibfield  {author} {\bibinfo {author} {\bibfnamefont {G.~F.}\ \bibnamefont
  {{Lindal}}},\ }\href {\doibase 10.1086/116119} {\bibfield  {journal}
  {\bibinfo  {journal} {\aj}\ }\textbf {\bibinfo {volume} {103}},\ \bibinfo
  {pages} {967} (\bibinfo {year} {1992})}\BibitemShut {NoStop}%
\bibitem [{\citenamefont {{Schinder}}\ \emph {et~al.}(2012)\citenamefont
  {{Schinder}}, \citenamefont {{Flasar}}, \citenamefont {{Marouf}},
  \citenamefont {{French}}, \citenamefont {{McGhee}}, \citenamefont {{Kliore}},
  \citenamefont {{Rappaport}}, \citenamefont {{Barbinis}}, \citenamefont
  {{Fleischman}},\ and\ \citenamefont {{Anabtawi}}}]{2012Icar..221.1020S}%
  \BibitemOpen
  \bibfield  {author} {\bibinfo {author} {\bibfnamefont {P.~J.}\ \bibnamefont
  {{Schinder}}}, \bibinfo {author} {\bibfnamefont {F.~M.}\ \bibnamefont
  {{Flasar}}}, \bibinfo {author} {\bibfnamefont {E.~A.}\ \bibnamefont
  {{Marouf}}}, \bibinfo {author} {\bibfnamefont {R.~G.}\ \bibnamefont
  {{French}}}, \bibinfo {author} {\bibfnamefont {C.~A.}\ \bibnamefont
  {{McGhee}}}, \bibinfo {author} {\bibfnamefont {A.~J.}\ \bibnamefont
  {{Kliore}}}, \bibinfo {author} {\bibfnamefont {N.~J.}\ \bibnamefont
  {{Rappaport}}}, \bibinfo {author} {\bibfnamefont {E.}~\bibnamefont
  {{Barbinis}}}, \bibinfo {author} {\bibfnamefont {D.}~\bibnamefont
  {{Fleischman}}}, \ and\ \bibinfo {author} {\bibfnamefont {A.}~\bibnamefont
  {{Anabtawi}}},\ }\href {\doibase 10.1016/j.icarus.2012.10.021} {\bibfield
  {journal} {\bibinfo  {journal} {\icarus}\ }\textbf {\bibinfo {volume}
  {221}},\ \bibinfo {pages} {1020} (\bibinfo {year} {2012})}\BibitemShut
  {NoStop}%
\bibitem [{\citenamefont {{Schinder}}\ \emph {et~al.}(2015)\citenamefont
  {{Schinder}}, \citenamefont {{Flasar}}, \citenamefont {{Marouf}},
  \citenamefont {{French}}, \citenamefont {{Anabtawi}}, \citenamefont
  {{Barbinis}},\ and\ \citenamefont {{Kliore}}}]{2015RaSc...50..712S}%
  \BibitemOpen
  \bibfield  {author} {\bibinfo {author} {\bibfnamefont {P.~J.}\ \bibnamefont
  {{Schinder}}}, \bibinfo {author} {\bibfnamefont {F.~M.}\ \bibnamefont
  {{Flasar}}}, \bibinfo {author} {\bibfnamefont {E.~A.}\ \bibnamefont
  {{Marouf}}}, \bibinfo {author} {\bibfnamefont {R.~G.}\ \bibnamefont
  {{French}}}, \bibinfo {author} {\bibfnamefont {A.}~\bibnamefont
  {{Anabtawi}}}, \bibinfo {author} {\bibfnamefont {E.}~\bibnamefont
  {{Barbinis}}}, \ and\ \bibinfo {author} {\bibfnamefont {A.~J.}\ \bibnamefont
  {{Kliore}}},\ }\href {\doibase 10.1002/2015RS005690} {\bibfield  {journal}
  {\bibinfo  {journal} {Radio Science}\ }\textbf {\bibinfo {volume} {50}},\
  \bibinfo {pages} {712} (\bibinfo {year} {2015})}\BibitemShut {NoStop}%
\bibitem [{\citenamefont {{Roques}}\ \emph {et~al.}(1994)\citenamefont
  {{Roques}}, \citenamefont {{Sicardy}}, \citenamefont {{French}},
  \citenamefont {{Hubbard}}, \citenamefont {{Barucci}}, \citenamefont
  {{Bouchet}}, \citenamefont {{Brahic}}, \citenamefont {{Gehrels}},
  \citenamefont {{Gehrels}}, \citenamefont {{Grenier}}, \citenamefont {{Le
  Bertre}}, \citenamefont {{Lecacheux}}, \citenamefont {{Maillard}},
  \citenamefont {{McLaren}}, \citenamefont {{Perrier}}, \citenamefont
  {{Vilas}},\ and\ \citenamefont {{Waterworth}}}]{1994A&A...288..985R}%
  \BibitemOpen
  \bibfield  {author} {\bibinfo {author} {\bibfnamefont {F.}~\bibnamefont
  {{Roques}}}, \bibinfo {author} {\bibfnamefont {B.}~\bibnamefont {{Sicardy}}},
  \bibinfo {author} {\bibfnamefont {R.~G.}\ \bibnamefont {{French}}}, \bibinfo
  {author} {\bibfnamefont {W.~B.}\ \bibnamefont {{Hubbard}}}, \bibinfo {author}
  {\bibfnamefont {A.}~\bibnamefont {{Barucci}}}, \bibinfo {author}
  {\bibfnamefont {P.}~\bibnamefont {{Bouchet}}}, \bibinfo {author}
  {\bibfnamefont {A.}~\bibnamefont {{Brahic}}}, \bibinfo {author}
  {\bibfnamefont {J.-A.}\ \bibnamefont {{Gehrels}}}, \bibinfo {author}
  {\bibfnamefont {T.}~\bibnamefont {{Gehrels}}}, \bibinfo {author}
  {\bibfnamefont {I.}~\bibnamefont {{Grenier}}}, \bibinfo {author}
  {\bibfnamefont {T.}~\bibnamefont {{Le Bertre}}}, \bibinfo {author}
  {\bibfnamefont {J.}~\bibnamefont {{Lecacheux}}}, \bibinfo {author}
  {\bibfnamefont {J.~P.}\ \bibnamefont {{Maillard}}}, \bibinfo {author}
  {\bibfnamefont {R.~A.}\ \bibnamefont {{McLaren}}}, \bibinfo {author}
  {\bibfnamefont {C.}~\bibnamefont {{Perrier}}}, \bibinfo {author}
  {\bibfnamefont {F.}~\bibnamefont {{Vilas}}}, \ and\ \bibinfo {author}
  {\bibfnamefont {M.~D.}\ \bibnamefont {{Waterworth}}},\ }\href
  {http://adsabs.harvard.edu/abs/1994A%26A...288..985R} {\bibfield  {journal}
  {\bibinfo  {journal} {\aap}\ }\textbf {\bibinfo {volume} {288}},\ \bibinfo
  {pages} {985} (\bibinfo {year} {1994})}\BibitemShut {NoStop}%
\bibitem [{\citenamefont {{Sicardy}}\ \emph {et~al.}(2006)\citenamefont
  {{Sicardy}}, \citenamefont {{Colas}}, \citenamefont {{Widemann}},
  \citenamefont {{Bellucci}}, \citenamefont {{Beisker}}, \citenamefont
  {{Kretlow}}, \citenamefont {{Ferri}}, \citenamefont {{Lacour}}, \citenamefont
  {{Lecacheux}}, \citenamefont {{Lellouch}}, \citenamefont {{Pau}},
  \citenamefont {{Renner}}, \citenamefont {{Roques}}, \citenamefont {{Fienga}},
  \citenamefont {{Etienne}}, \citenamefont {{Martinez}}, \citenamefont
  {{Glass}}, \citenamefont {{Baba}}, \citenamefont {{Nagayama}}, \citenamefont
  {{Nagata}}, \citenamefont {{Itting-Enke}}, \citenamefont {{Bath}},
  \citenamefont {{Bode}}, \citenamefont {{Bode}}, \citenamefont
  {{L{\"u}demann}}, \citenamefont {{L{\"u}demann}}, \citenamefont {{Neubauer}},
  \citenamefont {{Tegtmeier}}, \citenamefont {{Tegtmeier}}, \citenamefont
  {{Thom{\'e}}}, \citenamefont {{Hund}}, \citenamefont {{deWitt}},
  \citenamefont {{Fraser}}, \citenamefont {{Jansen}}, \citenamefont {{Jones}},
  \citenamefont {{Schoenau}}, \citenamefont {{Turk}}, \citenamefont
  {{Meintjies}}, \citenamefont {{Hernandez}}, \citenamefont {{Fiel}},
  \citenamefont {{Frappa}}, \citenamefont {{Peyrot}}, \citenamefont {{Teng}},
  \citenamefont {{Vignand}}, \citenamefont {{Hesler}}, \citenamefont {{Payet}},
  \citenamefont {{Howell}}, \citenamefont {{Kidger}}, \citenamefont {{Ortiz}},
  \citenamefont {{Naranjo}}, \citenamefont {{Rosenzweig}},\ and\ \citenamefont
  {{Rapaport}}}]{2006JGRE..11111S91S}%
  \BibitemOpen
  \bibfield  {author} {\bibinfo {author} {\bibfnamefont {B.}~\bibnamefont
  {{Sicardy}}}, \bibinfo {author} {\bibfnamefont {F.}~\bibnamefont {{Colas}}},
  \bibinfo {author} {\bibfnamefont {T.}~\bibnamefont {{Widemann}}}, \bibinfo
  {author} {\bibfnamefont {A.}~\bibnamefont {{Bellucci}}}, \bibinfo {author}
  {\bibfnamefont {W.}~\bibnamefont {{Beisker}}}, \bibinfo {author}
  {\bibfnamefont {M.}~\bibnamefont {{Kretlow}}}, \bibinfo {author}
  {\bibfnamefont {F.}~\bibnamefont {{Ferri}}}, \bibinfo {author} {\bibfnamefont
  {S.}~\bibnamefont {{Lacour}}}, \bibinfo {author} {\bibfnamefont
  {J.}~\bibnamefont {{Lecacheux}}}, \bibinfo {author} {\bibfnamefont
  {E.}~\bibnamefont {{Lellouch}}}, \bibinfo {author} {\bibfnamefont
  {S.}~\bibnamefont {{Pau}}}, \bibinfo {author} {\bibfnamefont
  {S.}~\bibnamefont {{Renner}}}, \bibinfo {author} {\bibfnamefont
  {F.}~\bibnamefont {{Roques}}}, \bibinfo {author} {\bibfnamefont
  {A.}~\bibnamefont {{Fienga}}}, \bibinfo {author} {\bibfnamefont
  {C.}~\bibnamefont {{Etienne}}}, \bibinfo {author} {\bibfnamefont
  {C.}~\bibnamefont {{Martinez}}}, \bibinfo {author} {\bibfnamefont {I.~S.}\
  \bibnamefont {{Glass}}}, \bibinfo {author} {\bibfnamefont {D.}~\bibnamefont
  {{Baba}}}, \bibinfo {author} {\bibfnamefont {T.}~\bibnamefont {{Nagayama}}},
  \bibinfo {author} {\bibfnamefont {T.}~\bibnamefont {{Nagata}}}, \bibinfo
  {author} {\bibfnamefont {S.}~\bibnamefont {{Itting-Enke}}}, \bibinfo {author}
  {\bibfnamefont {K.-L.}\ \bibnamefont {{Bath}}}, \bibinfo {author}
  {\bibfnamefont {H.-J.}\ \bibnamefont {{Bode}}}, \bibinfo {author}
  {\bibfnamefont {F.}~\bibnamefont {{Bode}}}, \bibinfo {author} {\bibfnamefont
  {H.}~\bibnamefont {{L{\"u}demann}}}, \bibinfo {author} {\bibfnamefont
  {J.}~\bibnamefont {{L{\"u}demann}}}, \bibinfo {author} {\bibfnamefont
  {D.}~\bibnamefont {{Neubauer}}}, \bibinfo {author} {\bibfnamefont
  {A.}~\bibnamefont {{Tegtmeier}}}, \bibinfo {author} {\bibfnamefont
  {C.}~\bibnamefont {{Tegtmeier}}}, \bibinfo {author} {\bibfnamefont
  {B.}~\bibnamefont {{Thom{\'e}}}}, \bibinfo {author} {\bibfnamefont
  {F.}~\bibnamefont {{Hund}}}, \bibinfo {author} {\bibfnamefont
  {C.}~\bibnamefont {{deWitt}}}, \bibinfo {author} {\bibfnamefont
  {B.}~\bibnamefont {{Fraser}}}, \bibinfo {author} {\bibfnamefont
  {A.}~\bibnamefont {{Jansen}}}, \bibinfo {author} {\bibfnamefont
  {T.}~\bibnamefont {{Jones}}}, \bibinfo {author} {\bibfnamefont
  {P.}~\bibnamefont {{Schoenau}}}, \bibinfo {author} {\bibfnamefont
  {C.}~\bibnamefont {{Turk}}}, \bibinfo {author} {\bibfnamefont
  {P.}~\bibnamefont {{Meintjies}}}, \bibinfo {author} {\bibfnamefont
  {M.}~\bibnamefont {{Hernandez}}}, \bibinfo {author} {\bibfnamefont
  {D.}~\bibnamefont {{Fiel}}}, \bibinfo {author} {\bibfnamefont
  {E.}~\bibnamefont {{Frappa}}}, \bibinfo {author} {\bibfnamefont
  {A.}~\bibnamefont {{Peyrot}}}, \bibinfo {author} {\bibfnamefont {J.~P.}\
  \bibnamefont {{Teng}}}, \bibinfo {author} {\bibfnamefont {M.}~\bibnamefont
  {{Vignand}}}, \bibinfo {author} {\bibfnamefont {G.}~\bibnamefont {{Hesler}}},
  \bibinfo {author} {\bibfnamefont {T.}~\bibnamefont {{Payet}}}, \bibinfo
  {author} {\bibfnamefont {R.~R.}\ \bibnamefont {{Howell}}}, \bibinfo {author}
  {\bibfnamefont {M.}~\bibnamefont {{Kidger}}}, \bibinfo {author}
  {\bibfnamefont {J.~L.}\ \bibnamefont {{Ortiz}}}, \bibinfo {author}
  {\bibfnamefont {O.}~\bibnamefont {{Naranjo}}}, \bibinfo {author}
  {\bibfnamefont {P.}~\bibnamefont {{Rosenzweig}}}, \ and\ \bibinfo {author}
  {\bibfnamefont {M.}~\bibnamefont {{Rapaport}}},\ }\href
  {http://adsabs.harvard.edu/abs/2006JGRE..11111S91S} {\bibfield  {journal}
  {\bibinfo  {journal} {Journal of Geophysical Research (Planets)}\ }\textbf
  {\bibinfo {volume} {111}},\ \bibinfo {eid} {E11S91} (\bibinfo {year}
  {2006})}\BibitemShut {NoStop}%
\bibitem [{\citenamefont {{Phinney}}\ and\ \citenamefont
  {{Anderson}}(1968)}]{1968JGR....73.1819P}%
  \BibitemOpen
  \bibfield  {author} {\bibinfo {author} {\bibfnamefont {R.~A.}\ \bibnamefont
  {{Phinney}}}\ and\ \bibinfo {author} {\bibfnamefont {D.~L.}\ \bibnamefont
  {{Anderson}}},\ }\href {\doibase 10.1029/JA073i005p01819} {\bibfield
  {journal} {\bibinfo  {journal} {\jgr}\ }\textbf {\bibinfo {volume} {73}},\
  \bibinfo {pages} {1819} (\bibinfo {year} {1968})}\BibitemShut {NoStop}%
\bibitem [{\citenamefont {{Steiner}}\ \emph {et~al.}(1999)\citenamefont
  {{Steiner}}, \citenamefont {{Kirchengast}},\ and\ \citenamefont
  {{Ladreiter}}}]{1999AnGeo..17..122S}%
  \BibitemOpen
  \bibfield  {author} {\bibinfo {author} {\bibfnamefont {A.~K.}\ \bibnamefont
  {{Steiner}}}, \bibinfo {author} {\bibfnamefont {G.}~\bibnamefont
  {{Kirchengast}}}, \ and\ \bibinfo {author} {\bibfnamefont {H.~P.}\
  \bibnamefont {{Ladreiter}}},\ }\href {\doibase 10.1007/s00585-999-0122-5}
  {\bibfield  {journal} {\bibinfo  {journal} {Annales Geophysicae}\ }\textbf
  {\bibinfo {volume} {17}},\ \bibinfo {pages} {122} (\bibinfo {year}
  {1999})}\BibitemShut {NoStop}%
\bibitem [{\citenamefont {{Bourgoin}}\ \emph {et~al.}(2019)\citenamefont
  {{Bourgoin}}, \citenamefont {{Zannoni}},\ and\ \citenamefont
  {{Tortora}}}]{2019A&A...624A..41B}%
  \BibitemOpen
  \bibfield  {author} {\bibinfo {author} {\bibfnamefont {A.}~\bibnamefont
  {{Bourgoin}}}, \bibinfo {author} {\bibfnamefont {M.}~\bibnamefont
  {{Zannoni}}}, \ and\ \bibinfo {author} {\bibfnamefont {P.}~\bibnamefont
  {{Tortora}}},\ }\href {\doibase 10.1051/0004-6361/201834962} {\bibfield
  {journal} {\bibinfo  {journal} {\aap}\ }\textbf {\bibinfo {volume} {624}},\
  \bibinfo {eid} {A41} (\bibinfo {year} {2019})}\BibitemShut {NoStop}%
\bibitem [{\citenamefont {{Quan}}(1957)}]{1957ArRMA...1...54Q}%
  \BibitemOpen
  \bibfield  {author} {\bibinfo {author} {\bibfnamefont {P.~M.}\ \bibnamefont
  {{Quan}}},\ }\href {\doibase 10.1007/BF00297996} {\bibfield  {journal}
  {\bibinfo  {journal} {Archive for Rational Mechanics and Analysis}\ }\textbf
  {\bibinfo {volume} {1}},\ \bibinfo {pages} {54} (\bibinfo {year}
  {1957})}\BibitemShut {NoStop}%
\bibitem [{\citenamefont {{Ehlers}}(1967)}]{1967ZNatA..22.1328E}%
  \BibitemOpen
  \bibfield  {author} {\bibinfo {author} {\bibfnamefont {J.}~\bibnamefont
  {{Ehlers}}},\ }\href {\doibase 10.1515/zna-1967-0906} {\bibfield  {journal}
  {\bibinfo  {journal} {Zeitschrift Naturforschung Teil A}\ }\textbf {\bibinfo
  {volume} {22}},\ \bibinfo {pages} {1328} (\bibinfo {year}
  {1967})}\BibitemShut {NoStop}%
\bibitem [{\citenamefont {Perlick}(2000)}]{perlick2000ray}%
  \BibitemOpen
  \bibfield  {author} {\bibinfo {author} {\bibfnamefont {V.}~\bibnamefont
  {Perlick}},\ }\href {https://books.google.it/books?id=tymnfkVOAJsC} {\emph
  {\bibinfo {title} {Ray Optics, Fermat’s Principle, and Applications to
  General Relativity}}},\ Lecture Notes in Physics Monographs\ (\bibinfo
  {publisher} {Springer-Verlag},\ \bibinfo {year} {2000})\BibitemShut {NoStop}%
\bibitem [{\citenamefont {Blanchet}\ \emph {et~al.}(2001)\citenamefont
  {Blanchet}, \citenamefont {Salomon}, \citenamefont {Teyssandier},\ and\
  \citenamefont {Wolf}}]{2001A&A...370..320B}%
  \BibitemOpen
  \bibfield  {author} {\bibinfo {author} {\bibfnamefont {L.}~\bibnamefont
  {Blanchet}}, \bibinfo {author} {\bibfnamefont {C.}~\bibnamefont {Salomon}},
  \bibinfo {author} {\bibfnamefont {P.}~\bibnamefont {Teyssandier}}, \ and\
  \bibinfo {author} {\bibfnamefont {P.}~\bibnamefont {Wolf}},\ }\href {\doibase
  10.1051/0004-6361:20010233} {\bibfield  {journal} {\bibinfo  {journal}
  {Astronomy and Astrophysics}\ }\textbf {\bibinfo {volume} {370}},\ \bibinfo
  {pages} {320} (\bibinfo {year} {2001})}\BibitemShut {NoStop}%
\bibitem [{\citenamefont {Hees}\ \emph {et~al.}(2012)\citenamefont {Hees},
  \citenamefont {Lamine}, \citenamefont {Reynaud}, \citenamefont {{Jaekel}},
  \citenamefont {{Le Poncin-Lafitte}}, \citenamefont {{Lainey}}, \citenamefont
  {{F{\"u}zfa}}, \citenamefont {{Courty}}, \citenamefont {{Dehant}},\ and\
  \citenamefont {{Wolf}}}]{2012CQGra..29w5027H}%
  \BibitemOpen
  \bibfield  {author} {\bibinfo {author} {\bibfnamefont {A.}~\bibnamefont
  {Hees}}, \bibinfo {author} {\bibfnamefont {B.}~\bibnamefont {Lamine}},
  \bibinfo {author} {\bibfnamefont {S.}~\bibnamefont {Reynaud}}, \bibinfo
  {author} {\bibfnamefont {M.-T.}\ \bibnamefont {{Jaekel}}}, \bibinfo {author}
  {\bibfnamefont {C.}~\bibnamefont {{Le Poncin-Lafitte}}}, \bibinfo {author}
  {\bibfnamefont {V.}~\bibnamefont {{Lainey}}}, \bibinfo {author}
  {\bibfnamefont {A.}~\bibnamefont {{F{\"u}zfa}}}, \bibinfo {author}
  {\bibfnamefont {J.-M.}\ \bibnamefont {{Courty}}}, \bibinfo {author}
  {\bibfnamefont {V.}~\bibnamefont {{Dehant}}}, \ and\ \bibinfo {author}
  {\bibfnamefont {P.}~\bibnamefont {{Wolf}}},\ }\href {\doibase
  10.1088/0264-9381/29/23/235027} {\bibfield  {journal} {\bibinfo  {journal}
  {Classical and Quantum Gravity}\ }\textbf {\bibinfo {volume} {29}},\ \bibinfo
  {pages} {235027} (\bibinfo {year} {2012})}\BibitemShut {NoStop}%
\bibitem [{\citenamefont {{Hees}}\ \emph {et~al.}(2014)\citenamefont {{Hees}},
  \citenamefont {{Bertone}},\ and\ \citenamefont {{Le
  Poncin-Lafitte}}}]{2014PhRvD..89f4045H}%
  \BibitemOpen
  \bibfield  {author} {\bibinfo {author} {\bibfnamefont {A.}~\bibnamefont
  {{Hees}}}, \bibinfo {author} {\bibfnamefont {S.}~\bibnamefont {{Bertone}}}, \
  and\ \bibinfo {author} {\bibfnamefont {C.}~\bibnamefont {{Le
  Poncin-Lafitte}}},\ }\href {\doibase 10.1103/PhysRevD.89.064045} {\bibfield
  {journal} {\bibinfo  {journal} {\prd}\ }\textbf {\bibinfo {volume} {89}},\
  \bibinfo {eid} {064045} (\bibinfo {year} {2014})}\BibitemShut {NoStop}%
\bibitem [{Note1()}]{Note1}%
  \BibitemOpen
  \bibinfo {note} {In this section, we use the convention that hated index
  starting from the first part of the Greek or Latin alphabet denote components
  expressed in the rotating frame.}\BibitemShut {Stop}%
\bibitem [{\citenamefont {Teyssandier}(2012)}]{2012CQGra..29x5010T}%
  \BibitemOpen
  \bibfield  {author} {\bibinfo {author} {\bibfnamefont {P.}~\bibnamefont
  {Teyssandier}},\ }\href {\doibase 10.1088/0264-9381/29/24/245010} {\bibfield
  {journal} {\bibinfo  {journal} {Classical and Quantum Gravity}\ }\textbf
  {\bibinfo {volume} {29}},\ \bibinfo {pages} {245010} (\bibinfo {year}
  {2012})}\BibitemShut {NoStop}%
\bibitem [{\citenamefont {Arfken}(1985)}]{garfken67math}%
  \BibitemOpen
  \bibfield  {author} {\bibinfo {author} {\bibfnamefont {G.}~\bibnamefont
  {Arfken}},\ }\href
  {https://www.bibsonomy.org/bibtex/206a7a290f700ffb0ec221872f672f5ae/drmatusek}
  {\emph {\bibinfo {title} {{Mathematical Methods for Physicists}}}},\ \bibinfo
  {edition} {3rd}\ ed.\ (\bibinfo  {publisher} {Academic Press, {Inc.}},\
  \bibinfo {address} {San Diego},\ \bibinfo {year} {1985})\BibitemShut
  {NoStop}%
\bibitem [{\citenamefont {{Born}}\ and\ \citenamefont
  {{Wolf}}(1999)}]{1999prop.book.....B}%
  \BibitemOpen
  \bibfield  {author} {\bibinfo {author} {\bibfnamefont {M.}~\bibnamefont
  {{Born}}}\ and\ \bibinfo {author} {\bibfnamefont {E.}~\bibnamefont
  {{Wolf}}},\ }\href {http://cdsads.u-strasbg.fr/abs/1999prop.book.....B}
  {\emph {\bibinfo {title} {Principles of Optics}}}\ (\bibinfo  {publisher}
  {Cambridge University Press, UK},\ \bibinfo {year} {1999})\ p.\ \bibinfo
  {pages} {986}\BibitemShut {NoStop}%
\bibitem [{\citenamefont {{Landau}}\ and\ \citenamefont
  {{Lifshitz}}(1960)}]{1960ecm..book.....L}%
  \BibitemOpen
  \bibfield  {author} {\bibinfo {author} {\bibfnamefont {L.~D.}\ \bibnamefont
  {{Landau}}}\ and\ \bibinfo {author} {\bibfnamefont {E.~M.}\ \bibnamefont
  {{Lifshitz}}},\ }\href
  {https://ui.adsabs.harvard.edu/abs/1960ecm..book.....L} {\emph {\bibinfo
  {title} {{Electrodynamics of continuous media}}}}\ (\bibinfo  {publisher}
  {Elsevier Science},\ \bibinfo {year} {1960})\BibitemShut {NoStop}%
\bibitem [{\citenamefont {{Poisson}}\ and\ \citenamefont
  {{Will}}(2014)}]{2014gravbookP}%
  \BibitemOpen
  \bibfield  {author} {\bibinfo {author} {\bibfnamefont {E.}~\bibnamefont
  {{Poisson}}}\ and\ \bibinfo {author} {\bibfnamefont {C.~M.}\ \bibnamefont
  {{Will}}},\ }\href {http://adsabs.harvard.edu/abs/2014grav.book.....P} {\emph
  {\bibinfo {title} {{Gravity}}}}\ (\bibinfo  {publisher} {Cambridge University
  Press},\ \bibinfo {year} {2014})\BibitemShut {NoStop}%
\bibitem [{\citenamefont {{Waite}}\ \emph {et~al.}(2012)\citenamefont
  {{Waite}}, \citenamefont {{Bell}}, \citenamefont {{Lorenz}}, \citenamefont
  {{Achterberg}},\ and\ \citenamefont {{Flasar}}}]{2012LPI....43.1232W}%
  \BibitemOpen
  \bibfield  {author} {\bibinfo {author} {\bibfnamefont {J.~H.}\ \bibnamefont
  {{Waite}}}, \bibinfo {author} {\bibfnamefont {J.~M.}\ \bibnamefont {{Bell}}},
  \bibinfo {author} {\bibfnamefont {R.}~\bibnamefont {{Lorenz}}}, \bibinfo
  {author} {\bibfnamefont {R.}~\bibnamefont {{Achterberg}}}, \ and\ \bibinfo
  {author} {\bibfnamefont {F.~M.}\ \bibnamefont {{Flasar}}},\ }in\ \href
  {https://ui.adsabs.harvard.edu/abs/2012LPI....43.1232W} {\emph {\bibinfo
  {booktitle} {Lunar and Planetary Science Conference}}},\ \bibinfo {series and
  number} {Lunar and Planetary Science Conference}\ (\bibinfo {year} {2012})\
  p.\ \bibinfo {pages} {1232}\BibitemShut {NoStop}%
\bibitem [{\citenamefont {{Press}}\ \emph {et~al.}(1992)\citenamefont
  {{Press}}, \citenamefont {{Teukolsky}}, \citenamefont {{Vetterling}},\ and\
  \citenamefont {{Flannery}}}]{1992nrfa.book.....P}%
  \BibitemOpen
  \bibfield  {author} {\bibinfo {author} {\bibfnamefont {W.~H.}\ \bibnamefont
  {{Press}}}, \bibinfo {author} {\bibfnamefont {S.~A.}\ \bibnamefont
  {{Teukolsky}}}, \bibinfo {author} {\bibfnamefont {W.~T.}\ \bibnamefont
  {{Vetterling}}}, \ and\ \bibinfo {author} {\bibfnamefont {B.~P.}\
  \bibnamefont {{Flannery}}},\ }\href
  {http://adsabs.harvard.edu/abs/1992nrfa.book.....P} {\emph {\bibinfo {title}
  {{Numerical recipes in FORTRAN. The art of scientific computing}}}}\
  (\bibinfo  {publisher} {Cambridge: University Press, 2nd ed.},\ \bibinfo
  {year} {1992})\BibitemShut {NoStop}%
\bibitem [{\citenamefont {{Linet}}\ and\ \citenamefont
  {{Teyssandier}}(2013)}]{2013CQGra..30q5008L}%
  \BibitemOpen
  \bibfield  {author} {\bibinfo {author} {\bibfnamefont {B.}~\bibnamefont
  {{Linet}}}\ and\ \bibinfo {author} {\bibfnamefont {P.}~\bibnamefont
  {{Teyssandier}}},\ }\href {\doibase 10.1088/0264-9381/30/17/175008}
  {\bibfield  {journal} {\bibinfo  {journal} {Classical and Quantum Gravity}\
  }\textbf {\bibinfo {volume} {30}},\ \bibinfo {eid} {175008} (\bibinfo {year}
  {2013})}\BibitemShut {NoStop}%
\bibitem [{\citenamefont {{Landau}}\ and\ \citenamefont
  {{Lifshitz}}(1969)}]{1969mech.book.....L}%
  \BibitemOpen
  \bibfield  {author} {\bibinfo {author} {\bibfnamefont {L.~D.}\ \bibnamefont
  {{Landau}}}\ and\ \bibinfo {author} {\bibfnamefont {E.~M.}\ \bibnamefont
  {{Lifshitz}}},\ }\href {http://adsabs.harvard.edu/abs/1969mech.book.....L}
  {\emph {\bibinfo {title} {Course of Theoretical Physics, Oxford: Pergamon
  Press, 1969, 2nd ed.}}}\ (\bibinfo  {publisher} {Elsevier Science},\ \bibinfo
  {year} {1969})\BibitemShut {NoStop}%
\bibitem [{\citenamefont {{Fjeldbo}}\ \emph {et~al.}(1971)\citenamefont
  {{Fjeldbo}}, \citenamefont {{Kliore}},\ and\ \citenamefont
  {{Eshleman}}}]{1971AJ.....76..123F}%
  \BibitemOpen
  \bibfield  {author} {\bibinfo {author} {\bibfnamefont {G.}~\bibnamefont
  {{Fjeldbo}}}, \bibinfo {author} {\bibfnamefont {A.~J.}\ \bibnamefont
  {{Kliore}}}, \ and\ \bibinfo {author} {\bibfnamefont {V.~R.}\ \bibnamefont
  {{Eshleman}}},\ }\href {\doibase 10.1086/111096} {\bibfield  {journal}
  {\bibinfo  {journal} {\aj}\ }\textbf {\bibinfo {volume} {76}},\ \bibinfo
  {pages} {123} (\bibinfo {year} {1971})}\BibitemShut {NoStop}%
\end{thebibliography}%

\newpage
\appendix

\section{Abel transform method}
\label{sec:Abel}

In this section, we show how the atmospheric time delay and the refractivity can both be directly inferred from real Doppler data by making use of an Abel transform method. The main novelty of the following approach relies on the fact of considering  the dragging of light due to the angular velocity of the rigid rotation of the atmosphere.

\subsection{Bending angle from frequency transfer}

A general expression for the frequency transfer can be inferred from \eqref{eq:qAqB} and \eqref{eq:lAlB} such as
\begin{equation}
  \frac{\nu_B}{\nu_A}=\left[\frac{\nu_B}{\nu_A}\right]_{\mathrm{vac}}\left(\cfrac{1+\displaystyle\sum_{m=1}^{\infty}(N_0)^m\cfrac{\underbar{$\bm l$}^{(m)}_B\cdot\bm\beta_B}{1+\bm\beta_B\cdot\underbar{$\bm l$}_B^{\mathrm{(vac)}}}}{1+\displaystyle\sum_{m=1}^{\infty}(N_0)^m\cfrac{\underbar{$\bm l$}^{(m)}_A\cdot\bm\beta_A}{1+\bm\beta_A\cdot\underbar{$\bm l$}_A^{\mathrm{(vac)}}}}\right)\text{,}
\end{equation}
where the first factor on the right-hand side represents the frequency transfer in a vacuum, namely
\begin{subequations}\label{eq:DopVac}
\begin{equation}
  \left[\frac{\nu_B}{\nu_A}\right]_{\mathrm{vac}}=\frac{(u^0)_B}{(u^0)_A}\left(\frac{1+\bm\beta_B\cdot\underbar{$\bm l$}_B^{\mathrm{(vac)}}}{1+\bm\beta_A\cdot\underbar{$\bm l$}_A^{\mathrm{(vac)}}}\right)\text{.}\label{eq:DopVac1}
\end{equation}
The two triples $\underbar{$\bm l$}_{A}^{\mathrm{(vac)}}$ and $\underbar{$\bm l$}_{B}^{\mathrm{(vac)}}$ represents the directions of the light ray in a vacuum at the emission and reception, that is to say
\begin{align}
  \underbar{$\bm l$}_{A}^{\mathrm{(vac)}}&=-\bm N_{AB}+O(G)\text{,}\\
  \underbar{$\bm l$}_{B}^{\mathrm{(vac)}}&=-\bm N_{AB}+O(G)\text{.}
\end{align}
\end{subequations}
The omitted terms, proportional to $G$, represent the gravitational effects that are neglected here for clarity. The reader is referred to \citet{2013CQGra..30q5008L} for a complete determination of the gravitational terms up to $G^3$ in the context of a static, spherically symmetric spacetime.

For applications in the Solar system we can always consider that the 3-velocities are small with respect to the speed of light in a vacuum, that is to say $\Vert\bm\beta_A\Vert$ and $\Vert\bm\beta_B\Vert\ll 1$. Thus, at first orders in $N_0$, $\Vert\bm\beta_A\Vert$, and $\Vert\bm\beta_B\Vert$, we get
\begin{equation}
  \frac{\nu_B}{\nu_A}=\left[\frac{\nu_B}{\nu_A}\right]_{\mathrm{vac}}\left(1+N_0\,\underbar{$\bm l$}^{(1)}_B\cdot\bm\beta_B-N_0\,\underbar{$\bm l$}^{(1)}_A\cdot\bm{\beta}_A\right)\text{.}
\end{equation}
After substituting for $\underbar{$\bm l$}^{(1)}_A$ and $\underbar{$\bm l$}^{(1)}_B$ from Eqs. \eqref{eq:lAlB1}, and making use of \eqref{eq:epsN0}, the previous equation now reads
\begin{equation}
  \frac{\nu_B}{\nu_A}=\left[\frac{\nu_B}{\nu_A}\right]_{\mathrm{vac}}\left(1+N_0\phi^{(1)}\bm n_K\cdot\bm{\beta}_{\mathrm{eff}}\right)\text{,}
  \label{eq:dopsimp}
\end{equation}
where we introduce $\bm\beta_{\mathrm{eff}}$, an effective velocity, such as
\begin{equation}
  \bm{\beta}_{\mathrm{eff}}=\left(\frac{\bm N_{AB}\cdot\bm x_B}{R_{AB}}\right)\bm\beta_A+\left(1-\frac{\bm N_{AB}\cdot\bm x_B}{R_{AB}}\right)\bm\beta_B\text{.}
\end{equation}

Let us notice that when the receiver is at infinity (like for one-way radio occultations by planets or satellites of the outer Solar system), we have
\begin{equation}
 \lim_{r_B\to\infty}\frac{\bm N_{AB}\cdot\bm x_B}{R_{AB}}=1\text{,}
\end{equation}
thus, the effective velocity reduces to
\begin{equation}
  \lim_{r_B\to\infty}\bm{\beta}_{\mathrm{eff}}=\bm\beta_A\text{.}
\end{equation}
On the other hand, if the emitter is at infinity (like for one-way stellar occultations by planets or satellites of the Solar system), we have
\begin{equation}
 \lim_{r_A\to\infty}\frac{\bm N_{AB}\cdot\bm x_B}{R_{AB}}=0\text{,}
\end{equation}
thus, the effective velocity reduces to
\begin{equation}
  \lim_{r_A\to\infty}\bm{\beta}_{\mathrm{eff}}=\bm\beta_B\text{.}
\end{equation}

Eq. \eqref{eq:dopsimp} allows one to determine the first order bending angle at each time step from real Doppler data. Because the impact parameter is only given by the geometry at a given time, the data eventually provides $N_0\phi^{(1)}(K,\mathcal{H})$. 

\subsection{Atmospheric time delay from bending angle}

The refractive delay function can then be straightforwardly retrieved from the bending angle since, according to \eqref{eq:epsN0}, the bending angle is the derivative of the delay function with respect to $K$.

At first post-Minkowskian order, we recall that the delay function and the bending angle (see Eqs.~\eqref{eq:delRrPMN0G} and \eqref{eq:epsPM}, respectively) are given by
\begin{equation}
  \Delta(K,\mathcal{H})=N_0\Delta^{(1)}(K,\mathcal{H})\text{,}
\end{equation}
and
\begin{equation}
  \phi(K,\mathcal H)=N_0\phi^{(1)}(K,\mathcal H)\text{.}
\end{equation}
Therefore, by making use of \eqref{eq:epsN0}, we find
\begin{equation}
  \Delta(K,\mathcal{H})=\int_{K}^{\mathcal{H}}\phi(K',\mathcal H)\dd K'\text{,}
  \label{eq:delben}
\end{equation}
where the constant of integration has been chosen such that
\begin{equation}
  \Delta(\mathcal{H},\mathcal{H})=0
\end{equation}
considering that the bending angle is null at the beginning of the occultation, that is to say
\begin{equation}
  \phi(\mathcal{H},\mathcal{H})=0\text{.}
  \label{eq:varepsIC}
\end{equation}

After determining $\phi(K,\mathcal{H})$ thanks to \eqref{eq:dopsimp}, Eq.~\eqref{eq:delben} allows one to determine the atmospheric time delay and then the total light time.

\subsection{Refractivity from bending angle}

We saw back in Sec. \ref{sec:TTFRO} that $\Delta^{(1)}(K,\mathcal{H})$ is defined by \eqref{eq:del1KHdef}. If one applies the following change of variables
\begin{subequations}
\begin{align}
  a&=K^2-\mathcal{H}^2\text{,}\\
  b&=r^2-\mathcal{H}^2\text{,}
\end{align}
\end{subequations}
one can rewrite \eqref{eq:del1KHdef} as
\begin{equation}
  \Delta^{(1)}(a)=-C^2\int_0^a\mathcal N(b)\frac{\dd b}{\sqrt{b-a}}\text{.}
\end{equation}

Interestingly, it can be seen that this expression is a special case of Abel transform (see e.g. \cite{1968JGR....73.1819P,1969mech.book.....L}), which allows us to write
\begin{equation}
  C^2\mathcal N(b)=\frac{1}{\pi}\int_0^b\frac{\partial\Delta^{(1)}}{\partial a}\frac{\dd a}{\sqrt{a-b}}\text{.}
  \label{eq:Nr}
\end{equation}
Going back to the previous set of variables, and making use of \eqref{eq:epsN0}, we infer the following relationship after multiplying both sides of \eqref{eq:Nr} by $N_0$
\begin{equation}
  C^2N(r)=\frac{1}{\pi}\int_{r}^{\mathcal{H}}\phi(K,\mathcal{H})\,\frac{\dd K}{\sqrt{K^2-r^2}}\text{.}
\end{equation}
After integrating by parts the right-hand-side we get
\begin{align}
  C^2N(r)&=\frac{1}{\pi}\int^{\phi(r,\mathcal{H})}_{0}\mathrm{argch}\left(\frac{K(\phi')}{r}\right)\dd\phi'\text{.}
  \label{eq:NAbel}
\end{align}
where we used \eqref{eq:varepsIC} to show that
\begin{equation}
  \left[\phi(K,\mathcal{H})\,\mathrm{argch}\left(\frac{K}{r}\right)\right]_{K=r}^{K=\mathcal H}=0\text{.}
\end{equation}
The expression for $C^{2}$ can be inferred from \eqref{eq:C2}.

It is interesting to confront \eqref{eq:NAbel} with the standard Abel transform \citep{1971AJ.....76..123F} which usually provides an expression as follows
\begin{equation}
  n(r)=\exp\left[\frac{1}{\pi}\int^{\phi(r,\mathcal{H})}_{0}\mathrm{argch}\left(\frac{K(\phi')}{r}\right)\dd\phi'\right]\text{.}
  \label{eq:nAbel}
\end{equation}
According to \eqref{eq:defN}, in the limiting case where $C^2\to 1$ (i.e. no light-dragging effect), it is seen that \eqref{eq:NAbel} corresponds to the first order expression of \eqref{eq:nAbel}. Thus, the novelty of \eqref{eq:NAbel} with respect to \eqref{eq:nAbel} consists in taking into account the light-dragging effect through the geometric factor $C^2$.

\end{document}